\documentclass[english,final]{IEEEtran}
\usepackage[T1]{fontenc}
\usepackage[latin9]{inputenc}
\usepackage{bm}
\usepackage{amsmath}
\usepackage{amssymb}
\usepackage{graphicx}

\makeatletter

\providecommand{\tabularnewline}{\\}

  \newtheorem{thm}{Theorem}
  \newtheorem{lemma}{Lemma}
  \newtheorem{cor}{Corollary}
  \newtheorem{remrk}{Remark}
  \newtheorem{definition}{Definition}

\usepackage{amsmath,amssymb}
\usepackage[level=0]{wgroup_message}  
\usepackage{color}  
\usepackage{graphicx,psfrag,cite,subfigure}

\usepackage{cite}

\author{
Junting~Chen, \textit{Member,~IEEE},
Haifan~Yin, \\
Laura~Cottatellucci, \textit{Member,~IEEE},
David~Gesbert, \textit{Fellow,~IEEE}

\thanks{This work was supported by Huawei France Research Center.}
\thanks{The authors are with the Department of Communication Systems, EURECOM, Sophia-Antipolis, France (Email:\{chenju, yin, cottatel, gesbert\}@eurecom.fr). }
}

\makeatletter
\renewcommand*{\@opargbegintheorem}[3]{\@IEEEtmpitemindent\itemindent\topsep 0pt\rmfamily \trivlist%
      \item[\hskip \labelsep{\indent\itshape #1\ #2}] \textit{(#3):}\ \itemindent\@IEEEtmpitemindent}
\makeatother

\usepackage[acronym]{glossaries}
\newcommand{\newac}{\newacronym}
\newcommand{\ac}{\gls}
\newcommand{\Ac}{\Gls}

\makeglossaries
\newac{speb}{SPEB}{square position error bound}
\newac[plural=EFIMs,firstplural=Fisher information matrices (EFIMs)]{efim}{EFIM}{Fisher information matrix}
\newac{ne}{NE}{Nash equilibrium}
\newac{mse}{MSE}{mean squared error}
\newac{toa}{TOA}{time-of-arrival}
\newac{snr}{SNR}{signal-to-noise ratio}
\newac{lan}{LAN}{local area network}
\newac{psd}{PSD}{positive semidefinite}
\newac{pd}{PD}{positive definite}
\newac{wrt}{w.r.t.}{with respect to}
\newac{lhs}{L.H.S.}{left hand side}
\newac{wp1}{w.p.1}{with probability 1}
\newac{kkt}{KKT}{Karush-Kuhn-Tucker}
\newac{wlog}{w.l.o.g.}{without loss of generality}
\newac{mle}{MLE}{maximum likelihood estimation}
\newac{gps}{GPS}{global positioning system}
\newac{rssi}{RSSI}{received signal strength}
\newac{mimo}{MIMO}{multiple-input multiple-output}
\newac{csi}{CSI}{channel state information}
\newac{fdd}{FDD}{frequency division multiplexing}
\newac{ms}{MS}{mobile station}
\newac{bs}{BS}{base station}
\newac{d2d}{D2D}{device-to-device}
\newac{slnr}{SLNR}{signal-to-interference-leakage-and-noise-ratio}
\newac{ula}{ULA}{uniform linear antenna array}
\newac{pas}{PAS}{power angular spectrum}
\newac{mmse}{MMSE}{minimum mean square error}
\newac{zf}{ZF}{zero-forcing}
\newac{rzf}{RZF}{regularized zero-forcing}
\newac{as}{AS}{angular spread}
\newac{aod}{AOD}{angle of departure}
\newac{iid}{i.i.d.}{independent and identically distributed} 
\newac{sinr}{SINR}{signal-to-interference-and-noise ratio}
\newac{tdd}{TDD}{time-division duplex}
\newac{rvq}{RVQ}{random vector quantization}
\newac{rhs}{R.H.S.}{right hand side}
\newac{mrc}{MRC}{maximum ratio combining}
\newac{cdf}{CDF}{cumulative distribution function}
\newac{a.s.}{a.s.}{almost surely}
\newac{los}{LOS}{line-of-sight}
\newac{jsdm}{JSDM}{joint spatial division and multiplexing}
\newac{map}{MAP}{maximum a posteriori}
\newac{klt}{KLT}{Karhunen-Lo\`eve Transform}
\newac{lbe}{LBE}{link bargaining equilibrium}
\newac{se}{SE}{Stackelberg equilibrium}
\newac{uav}{UAV}{unmanned aerial vehicle}
\newac{nlos}{NLOS}{non-line-of-sight}
\newac{pdf}{PDF}{probability density function}
\newac{em}{EM}{expectation-maximization}
\newac{knn}{KNN}{$k$-nearest neighbor}

\makeatother

\usepackage{babel}
\begin{document}

\title{Efficient Feedback Mechanisms for FDD Massive MIMO under User-level
Cooperation }

\maketitle
%
\setkeys{Gin}{width=1.0\columnwidth}

%
\begin{abstract}
\Ac{csi} feedback is a challenging issue in \ac{fdd} massive MIMO
systems. This paper studies a cooperative feedback scheme, where the
users first exchange their \ac{csi} with each other by exploiting
\ac{d2d} communications, then compute the precoder by themselves,
and feed back the precoder to the \ac{bs}. Analytical results are
derived to show that the cooperative precoder feedback is more efficient
than the \ac{csi} feedback in terms of interference mitigation. Under
the constraint of limited \ac{d2d} communication capacity, we develop
an adaptive \ac{csi} exchange strategy based on signal subspace projection
and optimal bit partition. Numerical results demonstrate that the
proposed cooperative precoder feedback scheme with adaptive CSI exchange
significantly outperforms the CSI feedback scheme, even when the \ac{csi}
is exchanged via rate-limited \ac{d2d} communications.
\end{abstract}
\begin{keywords}
Massive MIMO, device-to-device, limited feedback, precoder feedback,
subspace projection
\end{keywords}

\section{Introduction}

Massive \ac{mimo} is widely considered to be one of the key enabling
technologies for future wireless communication systems \cite{Rusek:2013dz,HoyTenDeb:J13,LarEdfTufMar:M14,HuhCaiPapHar:J12}.
With more antennas at the \ac{bs}, massive \ac{mimo} systems have
more degrees of freedom to exploit for spatial multiplexing and interference
suppression. However, realizing such performance gain requires additional
efforts on acquiring the \ac{csi}, which has a large dimension. A
number of works focus on \ac{tdd} systems, where channel reciprocity
can be exploited to obtain the downlink \ac{csi} from the uplink
pilots transmitted by the users \cite{marzetta2010noncooperative,JosAshMar:J11,yin2013coordinated}.
However, \ac{fdd} systems are still dominant in current cellular
networks \cite{ChoLovBid:J14,NohZolLov:J16}.

Conventional limited feedback schemes rely on pre-defined codebooks
to quantize and feedback the channel vector \cite{au2007performance,love2006limited,jindal2006mimo,yoo2007multi,bhagavatula2011adaptive,yuan2012bit}.
However, these methods are not scalable to massive MIMO, because the
size of the codebook is exponential to the number of feedback bits,
which should increase linearly with the number of transmit antennas
in order to realize the full multiplexing gain \cite{jindal2006mimo}.
Hence designing improved feedback schemes in the context of FDD massive
MIMO is both challenging and timely. Among the state-of-the-art feedback
schemes, trellis-coded quantizers were studied in \cite{ChoChaLov:J13,ChoLovKim:J15}
for massive MIMO with moderate to high feedback loading, using source
coding techniques with only a small codebook. In addition, compressive
sensing techniques were applied to channel estimation and feedback
in \cite{RaoLau:J14,GaoDaiWanChe:J15}, under the sparsity assumption
for the massive MIMO channel. In contrast to developing vector quantization
and reconstruction techniques for massive MIMO, a two-layer precoding
structure was introduced to relieve the burden of instantaneous CSI
feedback by exploiting the low rank property of the channel covariance
matrices \cite{Adhikary:2012vn,Adhikary:2013kx,Chen14-JSAC,KimLeeSun15}.
However, the low rank property may not exist in some propagation scenarios
due to possible rich scattering environment and sufficient antenna
spacing.

In this paper, we tackle the \ac{csi} limited feedback issue in FDD
massive MIMO systems by exploiting user-level cooperation. Specifically,
we exploit the synergy between massive MIMO and \ac{d2d} communications,
where users are configured to exchange the (quantized) instantaneous
\ac{csi} with each other via \ac{d2d}, and feed back the precoder
(rather than the CSI) to the BS.  The intuition is that, first, experience
and analysis have shown that feedback resources  for MIMO precoding
are better used to convey information directly in the precoder domain
rather than in the channel domain. This is because greater mismatch
may be brought in by computing the MIMO precoder from the quantized
\ac{csi} feedback as quantization errors propagate during channel
inversion \cite{LovHeaLauGes:J08}. Second, computing the precoder
at the user side is not possible in classical MIMO systems without
\ac{d2d}, but it is feasible when \ac{d2d} is exploited. In the
ideal case of perfect D2D, \ac{csi} exchange  allows the users to
obtain the global CSI, compute, and feed back the precoder to the
\ac{bs}. Significant throughput gain of such precoder feedback scheme
has been demonstrated in prior work \cite{yin2014enabling}. 

However, the following two major issues need to be addressed: (i)
\emph{Does the precoder feedback scheme work under imperfect CSI exchange?}
(ii) \emph{How to efficiently quantize and exchange the CSI via D2D?
}Note that the prior work \cite{yin2014enabling}  required a group
leader to compute the precoder for all the users, and  it made an
ideal assumption that the users can obtain perfect global CSI. Such
assumption is difficult to be realized in practice due to limited
D2D channel capacity and transmission latency. Some preliminary analytical
results under limited D2D channel capacity were given in  \cite{chen2015precoder},
but it is still not known how to efficiently exchange the CSI among
the users for better performance. 

\mysubsubnote{For the comment to "efficiently exchange": a language tutor once told me that it is better to put the adverb close to the verb. So "efficiently exchange" sounds better than "exchange ... efficiently".}

To address these challenging issues, we develop strategies and analytical
results for two application scenarios of the cooperative precoder
feedback scheme. In the first scenario, we consider the users have
uncorrelated channels with identical path loss, and we analyze the
performance  under limited CSI exchange. In the second scenario, we
consider the users have non-identical channel statistics, where the
users may experience different path loss or have different signal
subspaces. We propose a novel CSI exchange strategy and derive the
optimal bit partition over each D2D link to achieve the minimum interference
leakage for the proposed cooperative precoder feedback scheme. The
key intuition is that, the users only need to share the portion of
CSI that lies in the overlapping signal subspace. For example, in
the extreme case when two users have non-overlapping signal subspaces,
they do not need to exchange the CSI. In the other extreme case when
two users have identical signal subspace and identical path loss,
they need high quality CSI exchange. 

The major findings and contributions of this paper are summarized
as follows: 
\begin{itemize}
\item Under the user-level cooperative feedback framework, we propose a
novel CSI exchange strategy. Based on this, we developed optimal bit
partition algorithms for the CSI quantizers for each D2D link. 
\item We analyze the performance of the cooperative precoder feedback scheme
under limited rate for D2D CSI exchange. We found that the proposed
scheme can reduce the interference leakage to $1/(K-1)$ of the CSI
feedback scheme in a $K$-user system under uncorrelated MIMO channels
with identical path loss. 
\item We demonstrate that even with limited D2D capacity, the cooperative
precoder feedback scheme can significantly outperform the CSI feedback
scheme. Moreover, the proposed CSI exchange strategy with optimal
bit partition saves up to half of the bits for CSI exchange. 
\end{itemize}

The rest of the paper is organized as follows. Section \ref{sec:system-model}
introduces the precoder feedback scheme with the CSI exchange mechanism.
Section \ref{sec:interference-analysis} analyzes the interference
leakage under uncorrelated channels with identical path loss. Section
\ref{sec:CSI-exchange-design} studies the efficient CSI exchange
strategy under non-identical channels, where users experience different
signal subspaces and path loss. Numerical results are demonstrated
in Section \ref{sec:numerical} and conclusions are given in Section
\ref{sec:conclusion}.

\emph{Notations}: The notations $\|\mathbf{a}\|$ and $\|\mathbf{A}\|$
denote the Euclidean norm of vector $\mathbf{a}$ and the matrix $2$-norm
of $\mathbf{A}$, respectively. In addition, $(\cdot)^{\text{H}}$
denotes the Hermitian transpose and $\text{tr}\{\mathbf{A}\}$ denotes
the trace of matrix $\mathbf{A}$.

\section{System Model}

\label{sec:system-model}

In this section, we elaborate the system model for massive MIMO downlink
transmission and introduce the cooperative precoder feedback based
on user-level cooperation. 

\subsection{Signal Model}

Consider a single cell massive MIMO network, where the \ac{bs} equips
with $N_{t}$ antennas and serves $K$ users. Denote the downlink
channel of user $k$ as $\mathbf{h}_{k}^{\text{H}}$, where $\mathbf{h}_{k}\in\mathbb{C}^{N_{t}}$
is a column vector and is independent across users. The received signal
of user $k$ is given by 
\[
y_{k}=\sqrt{\frac{P}{K}}\mathbf{h}_{k}^{\text{H}}\mathbf{w}_{k}s_{k}+\sqrt{\frac{P}{K}}\sum_{j\neq k}\mathbf{h}_{k}^{\text{H}}\mathbf{w}_{j}s_{j}+n_{k}
\]
where $s_{k}$ is the transmitted symbol with $\mathbb{E}\{|s_{k}|^{2}\}=1$,
$\mathbf{w}_{k}\in\mathbb{C}^{N_{t}}$ is the precoder with $\|\mathbf{w}_{k}\|=1$,
$n_{k}\sim\mathcal{C}\mathcal{N}(0,1)$ is the additive Gaussian noise,
and $P$ is the total transmission power. 

Assume that $\mathbf{h}_{k}$ follows distribution $\mathcal{CN}(\mathbf{0},l_{k}\mathbf{R}_{k})$,
where the covariance matrix $\mathbf{R}_{k}$ is normalized to $\text{tr}\{\mathbf{R}_{k}\}=N_{t}$
and $l_{k}$ denotes the path loss. The statistics $\{l_{k},\mathbf{R}_{k}\}$
is assumed known by all the users. Perfect \ac{csi} $\mathbf{h}_{k}$
is assumed available at each user $k$. In addition, consider that
the users exploit reliable \ac{d2d} communication links for finite
rate \ac{csi} exchange with each other. The CSI exchange and the
feedback strategies are specified as follows.

\subsection{Cooperative Precoder Feedback based on CSI Exchange}

Consider the system is operated in \ac{fdd} mode and explicit feedback
is required for \ac{csi} acquisition and downlink precoding. Suppose
each user has $B_{\text{f}}$ bits for the feedback to the BS. In
conventional CSI feedback, each user quantizes the channel $\mathbf{h}_{k}$
into $\hat{\mathbf{h}}_{k}$ coded by $B_{\text{f}}$ bits and feeds
back $\hat{\mathbf{h}}_{k}$ to the BS. Based on the global CSI $\hat{\mathbf{H}}=[\hat{\mathbf{h}}_{1},\hat{\mathbf{h}}_{2},\dots,\hat{\mathbf{h}}_{K}]$,
the BS computes the precoding matrix $\mathbf{W}=[\mathbf{w}_{1},\mathbf{w}_{2},\dots,\mathbf{w}_{K}]$.

In this paper, we consider a \emph{cooperative feedback} scheme, which
consists of two phases:
\begin{itemize}
\item \emph{CSI Exchange:} Each user $k$ employs a quantizer $\mathcal{Q}_{kj}$
to share the quantized channel 
\[
\hat{\mathbf{h}}_{k}^{(j)}=\mathcal{Q}_{kj}(\mathbf{h}_{k})
\]
to user $j$ via D2D communication. After the CSI exchange, each user
$k$ knows the imperfect global CSI 
\begin{equation}
\hat{\mathbf{H}}_{k}=[\hat{\mathbf{h}}_{1}^{(k)},\,\hat{\mathbf{h}}_{2}^{(k)},\,\dots\hat{\mathbf{h}}_{k-1}^{(k)},\,\mathbf{h}_{k},\,\hat{\mathbf{h}}_{k+1}^{(k)},\,\dots,\,\hat{\mathbf{h}}_{K}^{(k)}].\label{eq:global-csi-at-user-k}
\end{equation}
\item \emph{Cooperative Feedback:} With the global CSI, each user first
computes the precoder 
\[
\mathbf{w}_{k}^{\text{c}}=\mathcal{W}_{k}(\hat{\mathbf{H}}_{k})
\]
and then feeds back the precoder $\mathbf{w}_{k}^{\text{c}}$ to the
BS using $B_{\text{f}}$ bits. 
\end{itemize}
The BS applies the precoding vectors $\mathbf{w}_{k}=\mathbf{w}_{k}^{\text{c}}$
for downlink transmission. Fig. \ref{fig:rate-snr} illustrates a
signaling example of the cooperative precoder feedback scheme in two-user
case. 

\begin{figure}
\begin{centering}
\psfragscanon 
\psfrag{h1}[][][0.8]{$\mathbf{h}_1$} 
\psfrag{h2}[][][0.8]{$\mathbf{h}_2$} 
\psfrag{w1}[][][0.8]{$\mathbf{w}_1^{\text{c}}$} 
\psfrag{w2}[][][0.8]{$\mathbf{w}_2^{\text{c}}$} 
\psfrag{h10}[][][0.8]{$\hat{\mathbf{h}}_1^{(2)}$} 
\psfrag{h20}[][][0.8]{$\hat{\mathbf{h}}_2^{(1)}$}
\psfrag{U1}[][][0.8]{User 1}
\psfrag{U2}[][][0.8]{User 2}\includegraphics[width=0.5\columnwidth]{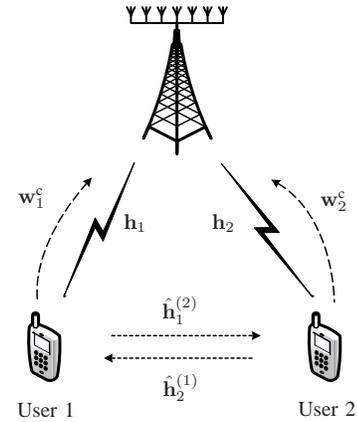}
\par\end{centering}
\caption{\label{fig:individual-precoding} A signaling example of the cooperative
precoder feedback scheme in two-user case.}
\end{figure}

Note that, if \ac{mrc} is considered as the precoding criterion for
$\mathcal{W}_{k}$, then there is no difference between precoder feedback
$\mathbf{w}_{k}^{\text{c}}$ and CSI feedback $\hat{\mathbf{h}}_{k}$.
By contrast, if \ac{zf} type criteria are used and the D2D CSI exchange
has a much higher rate than the feedback to the BS, then the cooperative
precoder feedback scheme $\mathcal{W}_{k}$ can exploit the advantage
of both knowing the self channel $\mathbf{h}_{k}$ perfectly and knowing
the channels from the other users more precisely. Such intuition will
be analyzed in the next section.

\section{Analysis of Cooperative Feedback for \\ Identically Uncorrelated
Channels}

\label{sec:interference-analysis}

In this section, we focus on identically uncorrelated channels, where
$l_{k}=1$ and $\mathbf{R}_{k}=\mathbf{I}$; i.e., the entries of
the channel vectors $\mathbf{h}_{k}$ are \ac{iid} and follow $\mathcal{C}\mathcal{N}(0,1)$.
Since the users have identical CSI statistics, the same quantizer
with the same rate $B_{\text{c}}$ can be used for the CSI exchange
in the proposed scheme. We develop analytical results to compare the
cooperative feedback scheme with the conventional CSI feedback scheme. 

\subsection{The Schemes}

The conventional CSI feedback scheme and the proposed cooperative
precoder feedback scheme are specified as follows. 

\emph{CSI feedback scheme:} \Ac{rvq} is used for channel quantization
and feedback, where each user $k$ has a \emph{channel codebook }$\mathcal{C}_{k}$
that contains $2^{B_{\text{f}}}$ $N_{t}$-dimensional unit norm isotropic
distributed vectors, and the channel $\mathbf{h}_{k}$ is quantized
as $\hat{\mathbf{g}}_{k}=\arg\max_{\mathbf{u}\in\mathcal{C}_{k}}|\mathbf{h}_{k}^{\text{H}}\mathbf{u}|$
and fed back to the BS. The \ac{bs} computes the precoder $\mathbf{w}_{k}$
for user $k$ as the normalized $k$th column of the precoding matrix
\begin{equation}
\mathbf{W}=\hat{\mathbf{G}}(\hat{\mathbf{G}}^{\text{H}}\hat{\mathbf{G}})^{-1}\label{eq:ZF}
\end{equation}
where $\hat{\mathbf{G}}$ is a $N_{t}\times K$ matrix with the $k$th
column given by the quantized channel $\hat{\mathbf{g}}_{k}$. Note
that, the channel magnitude $\|\mathbf{h}_{k}\|$ is not known by
the BS.

\emph{Cooperative precoder feedback scheme:} Each user has two codebooks:
the \emph{channel codebook }$\mathcal{C}_{kj}^{\text{c}}=\mathcal{C}_{k}^{\text{c}}$,
$j\neq k$, that contains $2^{B_{\text{c}}}$ $N_{t}$-dimensional
unit norm isotropic distributed vectors for the CSI exchange, and
the \emph{precoder codebook }$\mathcal{C}_{k}^{\text{w}}$ that contains
$2^{B_{\text{f}}}$ $N_{t}$-dimensional unit norm isotropic distributed
vectors for the feedback to the BS. In the CSI exchange phase, the
vector $\hat{\mathbf{h}}_{k}^{(j)}=\hat{\mathbf{h}}_{k}=\|\mathbf{h}_{k}\|\hat{\mathbf{g}}_{k}^{\text{c}}$
is shared to all the users $j\neq k$, where $\hat{\mathbf{g}}_{k}^{\text{c}}=\arg\max_{\mathbf{u}\in\mathcal{C}_{k}^{\text{c}}}|\mathbf{h}_{k}^{\text{H}}\mathbf{u}|$.\footnote{The channel magnitude $\|\mathbf{h}_{k}\|$ is assumed to be shared
among users with negligible distortion under additional $B_{\text{c}}^{(0)}$
bits. Note that $B_{\text{c}}^{(0)}$ needs not scales with $N_{t}$
and will not affect the main insights of the results, and hence is
ignored in this paper. } In the cooperative feedback phase, the vector that minimizes the
interference leakage is fed back to the BS (as the counterpart of
the ZF (\ref{eq:ZF}) in the CSI feedback scheme): 
\begin{equation}
\mathbf{w}_{k}^{\text{c}}=\arg\min_{\mathbf{w}\in\mathcal{C}_{k}^{\text{w}}}\sum_{j\neq k}|\hat{\mathbf{h}}_{j}^{\text{H}}\mathbf{w}|^{2}.\label{eq:minimum-interference-leakage-precoding}
\end{equation}

This section analyzes the performance in terms of interference leakage
defined as $I_{k}=\rho\sum_{j\neq k}|\mathbf{h}_{j}^{\text{H}}\mathbf{w}_{k}|^{2}$,
where $\rho=P/K$ denotes the power allocation. Before going through
the derivations, we first state the main result of this section as
follows. 
\begin{thm}
[Interference Leakage Upper Bound] Under large $N_{t}$ and $B_{\text{f}}$,
the average interference leakage under the precoder feedback scheme
is roughly upper bounded by 
\begin{equation}
\mathbb{E}_{\mathcal{H},\mathcal{C}}\left\{ I_{k}\right\} \lesssim\rho2^{-\frac{B_{\text{f}}}{K-1}}+\rho(K-1)2^{-\frac{B_{\text{c}}}{N_{t}-1}}\label{eq:interf-leakage-upper-bound}
\end{equation}
where the expectation $\mathbb{E}_{\mathcal{H},\mathcal{C}}\{\cdot\}$
is taken over the distributions of the channels and the codebooks.
\end{thm}

It is known that the interference leakage of the CSI feedback scheme
is lower bounded by $\mathbb{E}_{\mathcal{H},\mathcal{C}}\left\{ I_{k}\right\} >\rho(K-1)2^{-\frac{B_{\text{f}}}{N_{t}-1}}$
from \cite{jindal2006mimo}. Our result thus shows that for sufficiently
large $B_{\text{c}}$ for CSI exchange, the interference leakage from
the precoder feedback scheme is dominated by the first term of (\ref{eq:interf-leakage-upper-bound}),
which is $K-1$ times lower than the CSI feedback scheme and decreases
faster as $B_{\text{f}}$ increases.

\subsection{Characterization of the Interference Leakage}

We first characterize the interference leakage in terms of the precoding
vectors and the quantization errors. Without loss of generality, we
focus on the performance of user 1.
\begin{lemma}
[Characterization of the Interference Leakage]\label{lem:characterization-interference-leakage}
The mean of the interference leakage $I_{1}=\rho\sum_{j\neq1}|\mathbf{h}_{j}^{\text{H}}\mathbf{w}_{1}|^{2}$
can be characterized as 
\begin{equation}
\mathbb{E}_{\mathcal{H},\mathcal{C}}\left\{ I_{1}\right\} =\rho N_{t}\sum_{j\neq1}\mathbb{E}_{\mathcal{H},\mathcal{C}}\left\{ (1-Z_{j})\big|\hat{\mathbf{g}}_{j}^{\text{H}}\mathbf{w}_{1}\big|^{2}+Z_{j}\big|\mathbf{s}_{j}^{\text{H}}\mathbf{w}_{1}\big|^{2}\right\} \label{eq:Interference-leakage-characterization}
\end{equation}
where $Z_{j}\triangleq1-|\hat{\mathbf{g}}_{j}^{\text{H}}\mathbf{g}_{j}|^{2}$
with $\mathbf{g}_{j}=\mathbf{h}_{j}/\|\mathbf{h}_{j}\|$ and $\mathbf{s}_{j}\triangleq(\mathbf{I}-\hat{\mathbf{g}}_{j}\hat{\mathbf{g}}_{j}^{\text{H}})\mathbf{g}_{j}/\sqrt{Z_{j}}$.
\end{lemma}

\begin{IEEEproof}
Please refer to Appendix \ref{sec:app-proof-lem-IL-d2d-precoding}
for the proof.
\end{IEEEproof}

Note that the quantity $Z_{j}$ captures the channel quantization
error in magnitude, and $\mathbf{s}_{j}$ captures the difference
in direction between the quantized vector $\hat{\mathbf{g}}_{j}$
and the true channel $\mathbf{g}_{j}$ in the $(N_{t}-1)$-dimensional
space.\footnote{One can verify that $\mathbf{s}_{j}$ has unit norm and is orthogonal
to $\hat{\mathbf{g}}_{j}$. } Thus, Lemma \ref{lem:characterization-interference-leakage} illustrates
that the average interference is a sum of the interference leakage
due to precoding and the residual interference due to channel quantization
errors. 

Intuitive comparisons between precoder feedback and CSI feedback can
be made from (\ref{eq:Interference-leakage-characterization}). In
the CSI feedback scheme, the first term in (\ref{eq:Interference-leakage-characterization})
gives $\big|\hat{\mathbf{g}}_{j}^{\text{H}}\mathbf{w}_{1}\big|^{2}=0$
due to the \ac{zf} precoding at \ac{bs}. The second term characterizes
the interference leakage due to quantization error, which is in terms
of $B_{\text{f}}$. The results in \cite{jindal2006mimo} show that
it is roughly $N_{t}\mathbb{E}_{\mathcal{H},\mathcal{C}}\left\{ Z_{j}\big|\mathbf{s}_{j}^{\text{H}}\mathbf{w}_{1}\big|^{2}\right\} \approx2^{-\frac{B_{\text{f}}}{N_{t}-1}}$.

In the precoder feedback scheme, the first term $\big|\hat{\mathbf{g}}_{j}^{\text{H}}\mathbf{w}_{1}^{\text{c}}\big|^{2}\neq0$,
since $\mathbf{w}_{1}^{\text{c}}\in\mathcal{C}_{1}^{\text{w}}$ is
chosen from a finite number of vectors. By contrast, the second term
is affected by the quantization error in terms of $B_{\text{c}}$
for \ac{csi} exchange. Usually $B_{\text{c}}$ is large, and can
be evaluated using existing results. Specifically, the channel quantization
error bounds can be given as \cite{jindal2006mimo}

\begin{equation}
\frac{N_{t}-1}{N_{t}}2^{-\frac{B_{\text{c }}}{N_{t}-1}}<\mathbb{E}_{\mathcal{H},\mathcal{C}}\left\{ Z_{j}\right\} <2^{-\frac{B_{\text{c}}}{N_{t}-1}}\label{eq:Zj-chanenl-quantization-error-bounds}
\end{equation}
and $\mathbb{E}_{\mathcal{H},\mathcal{C}}\{\big|\mathbf{s}_{j}^{\text{H}}\mathbf{w}_{1}^{\text{c}}\big|^{2}\}=1/(N_{t}-1)$.
Moreover, $\big|\mathbf{s}_{j}^{\text{H}}\mathbf{w}_{1}\big|^{2}$
is independent of $Z_{j}$.\footnote{To see these results, note that $\mathbf{s}_{j}$ follows isotropic
distribution on the sphere in $(N_{t}-1)$-dimensional space, since
both the vectors $\hat{\mathbf{g}}_{k}^{\text{c}}$ in the channel
codebook $\mathcal{C}_{k}^{\text{c}}$ and the channel direction $\mathbf{g}_{k}$
are isotropically distributed in the $N_{t}$-dimensional space. As
a result, $\big|\mathbf{s}_{j}^{\text{H}}\mathbf{w}_{k}^{\text{c}}\big|^{2}$
follows a beta distribution $\mathcal{B}(1,N_{t}-2)$ for any unit
norm vector $\mathbf{w}_{k}^{\text{c}}$ as studied in \cite{jindal2006mimo},
and hence the mean is given by $1/(N_{t}-1)$.} As a result, $N_{t}\mathbb{E}_{\mathcal{H},\mathcal{C}}\{Z_{j}\big|\mathbf{s}_{j}^{\text{H}}\mathbf{w}_{1}\big|^{2}\}<\frac{N_{t}}{N_{t}-1}2^{-B_{\text{c}}/(N_{t}-1)}$.

In the following part, we focus on quantifying the first term in (\ref{eq:Interference-leakage-characterization})
for the interference leakage under the precoder feedback scheme with
user cooperation.

\subsection{Interference Upper Bound in a Two-user Case}

In two-user case, the precoder $\mathbf{w}_{1}$ only depends on $\hat{\mathbf{g}}_{2}$,
and the interference leakage from user $1$ is just the interference
at user $2$. Using this insight, the interference upper bound under
the precoder feedback scheme can be derived in the following theorem.
\begin{thm}
[Interference Upper Bound for Two Users]\label{thm:interference-leakage-two-user}
The mean of the interference leakage $I_{1}^{\text{c}}=\rho|\mathbf{h}_{2}^{\text{H}}\mathbf{w}_{1}^{\text{c}}|^{2}$
is upper bounded by 
\[
\mathbb{E}_{\mathcal{H},\mathcal{C}}\left\{ I_{1}^{\text{c}}\right\} \leq\frac{\rho N_{t}}{N_{t}-1}\left[2^{-B_{\text{f}}}+\bigg(1-\frac{N_{t}-1}{N_{t}}2^{-B_{\text{f}}}\bigg)2^{-\frac{B_{\text{c}}}{N_{t}-1}}\right].
\]
\end{thm}

\begin{IEEEproof}
Please refer to Appendix \ref{sec:app-proof-thm-intereference-two-user}
for the proof.
\end{IEEEproof}

The following corollary characterizes the case under perfect \ac{csi}
exchange among users.
\begin{cor}
[Interference Upper Bound under Perfect CSI Exchange]\label{cor:d2d-perfect-csi}
With perfect \ac{csi} exchange, i.e., $B_{\text{c}}=\infty$, the
interference is upper bounded by $\mathbb{E}_{\mathcal{H},\mathcal{C}}\left\{ I_{1}^{\text{c}}\right\} \leq\frac{\rho N_{t}}{N_{t}-1}2^{-B_{\text{f}}}$. 
\end{cor}

%


As a comparison, the mean of the interference $I_{1}=\rho|\mathbf{h}_{2}^{\text{H}}\mathbf{w}_{1}|^{2}$
under the CSI feedback scheme with \ac{zf} precoding at the \ac{bs}
can be bounded as \cite{jindal2006mimo}: 
\begin{equation}
\rho2^{-\frac{B_{\text{f}}}{N_{t}-1}}<\mathbb{E}_{\mathcal{H},\mathcal{C}}\left\{ I_{1}\right\} <\frac{\rho N_{t}}{N_{t}-1}2^{-\frac{B_{\text{f}}}{N_{t}-1}}.\label{eq:IL-BS-precoding}
\end{equation}

With imperfect \ac{csi} exchange, Theorem \ref{thm:interference-leakage-two-user}
shows that for $B_{\text{c}}\gg B_{\text{f}}$, the interference under
the precoder feedback scheme is smaller, and it decreases faster than
the CSI feedback scheme when increasing the number of feedback bits
$B_{\text{f}}$. On the other hand, when $B_{\text{c}}$ is small,
the interference under precoder feedback scheme is dominated by the
residual interference due to channel quantization errors for \ac{csi}
exchange among users.

\subsection{Interference Leakage in the $K$-user Case}

In $K>2$ user case, the precoding vector $\mathbf{w}_{1}^{\text{c}}$
depends on more than one channel vectors, and hence the exact distribution
of $\sum_{j\neq1}|\hat{\mathbf{h}}_{j}^{\text{H}}\mathbf{w}_{1}^{\text{c}}|^{2}$
is difficult to obtain. We resolve this challenge by using large system
approximations and the extreme value theory, assuming both $N_{t}$
and $2^{B_{\text{f}}}$ are large. 

Specifically, given a quantized channel realization $\{\hat{\mathbf{h}}_{j}\}_{j\neq1}$
and a sequence of \ac{iid} unit norm isotropic random vectors $\mathbf{w}_{1}^{\text{c}},\mathbf{w}_{2}^{\text{c}},\dots$
independent of $\{\hat{\mathbf{h}}_{j}\}_{j\neq1}$, we first approximate
the random variables $\widetilde{Y}_{i}\triangleq\sum_{j\neq1}|\hat{\mathbf{h}}_{j}^{\text{H}}\mathbf{w}_{i}^{\text{c}}|^{2}$
as independent chi-square random variables (multiplied by a scale
factor $\frac{1}{2}$) with degrees of freedom $2(K-1)$. Note that
such approximation becomes exact in large $N_{t}$. 

The following lemma gives the asymptotic distribution of $\widetilde{Y}_{i}$
under the large $N_{t}$ regime. 
\begin{lemma}
[Asymptotic Chi-square Distribution]\label{lem:Asymptotic-chi-square-approx}
Let $X_{1},X_{2},\dots,X_{N}$ be a sequence of \ac{iid} random variables
that follows chi-square distribution $\chi^{2}(2(K-1))$. Then, $(\widetilde{Y}_{1},\widetilde{Y}_{2},\dots,\widetilde{Y}_{N})$
converges to $\frac{1}{2}(X_{1},X_{2},\dots,X_{N})$ in distribution,
as $N_{t}\to\infty$.
\end{lemma}

\begin{IEEEproof}
Please refer to Appendix \ref{sec:app-proof-lem-asymptotic-chi-square-approx}
for the proof.
\end{IEEEproof}

Lemma \ref{lem:Asymptotic-chi-square-approx} shows that as $N_{t}$
becomes large, the variables $\widetilde{Y}_{i}$ and $\widetilde{Y}_{j}$
tend to become independent and $\frac{1}{2}\chi^{2}(2(K-1))$ chi-square
distributed. 

Consider the minimum interference leakage precoding criterion in (\ref{eq:minimum-interference-leakage-precoding}),
and note that the precoding vector $\mathbf{w}_{k}^{\text{c}}$ is
chosen from a set of \ac{iid} isotropic vectors in $\mathcal{C}_{k}^{\text{w}}$.
Thus, the resultant interference leakage $\sum_{j\neq1}|\hat{\mathbf{h}}_{j}^{\text{H}}\mathbf{w}_{k}^{\text{c}}|^{2}$
is approximately the minimum of $2^{B_{\text{f}}}$ \ac{iid} chi-square
distributed (with a constant factor $\frac{1}{2}$) random variables
$\widetilde{Y}_{i}=\sum_{j\neq1}|\hat{\mathbf{h}}_{j}^{\text{H}}\mathbf{w}_{i}^{\text{c}}|^{2}$,
$i=1,2,\dots,2^{B_{\text{f}}}$. As the codebook size $N=2^{B_{\text{f}}}$
is usually very large, one can apply extreme value theory to approximate
the distribution of $\min_{i}\widetilde{Y}_{i}$ in order to yield
simple expressions. 

Let $\hat{I}_{k}\triangleq\sum_{j\neq k}|\hat{\mathbf{h}}_{j}^{\text{H}}\mathbf{w}_{k}^{\text{c}}|^{2}$,
where $\mathbf{w}_{k}^{\text{c}}$ is chosen from the precoder codebook
$\mathcal{C}_{k}^{\text{w}}$ under minimum interference leakage criterion
(\ref{eq:minimum-interference-leakage-precoding}). Thus, $\hat{I}_{k}=\min_{i}\widetilde{Y}_{i}$.
Let $N=|\mathcal{C}_{k}^{\text{w}}|$. The asymptotic property of
$\hat{I}_{k}$ can be characterized in the following lemma. 
\begin{lemma}
[Asymptotic Distribution of $\hat{I}_{k}$]\label{lem:asymptotic-distribution}
The distribution of $\hat{I}_{k}$ satisfies
\[
\lim_{N\to\infty}\lim_{N_{t}\to\infty}\mathbb{P}\left\{ \hat{I}_{k}<\phi_{N}y\right\} =1-\exp(-y^{K-1}),\qquad x\geq0
\]
where 
\begin{equation}
\phi_{N}=\sup\left\{ x:\;\frac{1}{\Gamma(K-1)}\int_{0}^{x}t^{K-2}e^{-t}dt\leq\frac{1}{N}\right\} \label{eq:phi_N}
\end{equation}
in which $\Gamma(x)$ denotes the Gamma function. Moreover, for small
$K$, $\phi_{N}$ can be approximated by 
\begin{equation}
\phi_{N}\approx\Gamma(K)^{-\frac{1}{K-1}}N^{-\frac{1}{K-1}}.\label{eq:phi_N_approx}
\end{equation}
\end{lemma}

\begin{IEEEproof}
Please refer to Appendix \ref{sec:app-proof-lem-asymp-dist} for the
proof.
\end{IEEEproof}

Lemma \ref{lem:asymptotic-distribution} suggests that for large $N=2^{B_{\text{f}}}$
and large $N_{t}$, the interference leakage $\hat{I}_{k}$ due to
finite precoding can be approximated by a random variable $\phi_{N}W_{K-1}$
in distribution, where $W_{K-1}$ is Weibull distributed with \ac{cdf}
given by $f_{W}(x;K-1)=1-\exp(-x^{K-1})$, $x\geq0$, and mean $\mathbb{E}\{W_{K-1}\}=\Gamma\left(\frac{K}{K-1}\right)$.

With these results, the mean interference leakage under the precoder
feedback scheme can be derived in the following theorem.
\begin{thm}
[Interference Leakage for $K$ Users]\label{thm:Inteference-Leakage-K-users}
The mean of the interference leakage $I_{k}^{\text{c}}=\rho\sum_{j\neq k}|\mathbf{h}_{j}^{\text{H}}\mathbf{w}_{k}^{\text{c}}|^{2}$
under $K$-user networks can be approximated by 
\begin{align}
 & \mathbb{E}_{\mathcal{H},\mathcal{C}}\left\{ I_{k}^{\text{c}}\right\} \approx\rho\Gamma\left(\frac{K}{K-1}\right)\phi_{N}\nonumber \\
 & \qquad+\rho\left[\frac{N_{t}(K-1)}{N_{t}-1}-\frac{N_{t}-1}{N_{t}}\Gamma\left(\frac{K}{K-1}\right)\phi_{N}\right]2^{-\frac{B_{\text{c}}}{N_{t}-1}}\label{eq:interference-leakage-K-user}
\end{align}
where $\phi_{N}$ is given in (\ref{eq:phi_N}) with $N=2^{B_{\text{f}}}$.
In addition, for small $K$, 
\begin{align}
 & \mathbb{E}_{\mathcal{H},\mathcal{C}}\left\{ I_{k}^{\text{c}}\right\} \approx\rho\Phi(K)2^{-\frac{B_{\text{f}}}{K-1}}\nonumber \\
 & \qquad\qquad+\rho\left[\frac{N_{t}(K-1)}{N_{t}-1}-\frac{N_{t}-1}{N_{t}}\Phi(K)2^{-\frac{B_{\text{f}}}{K-1}}\right]2^{-\frac{B_{\text{c}}}{N_{t}-1}}\label{eq:interference-leakage-K-user-approx}
\end{align}
where $\Phi(K)\triangleq\Gamma(\frac{K}{K-1})\Gamma(K)^{-\frac{1}{K-1}}$.
\end{thm}

\begin{IEEEproof}
Using the results in Lemma \ref{lem:characterization-interference-leakage}
and \ref{lem:asymptotic-distribution}, the derivation is similar
to Theorem \ref{thm:interference-leakage-two-user}, and is omitted
here due to limited space. 
\end{IEEEproof}

One can numerically verify that the term $\Phi(K)$ is decreasing
in $K$ and $\Phi(K)\leq1$ for $K\geq2$. Therefore, under sufficiently
large $B_{\text{c}}$ and $N_{t}$, the interference leakage $\mathbb{E}_{\mathcal{H},\mathcal{C}}\left\{ I_{k}^{\text{c}}\right\} $
is roughly upper bounded by $\rho2^{-\frac{B_{\text{f}}}{K-1}}+\rho(K-1)2^{-\frac{B_{\text{c}}}{N_{t}-1}}$,
which is significantly smaller than that of the CSI feedback scheme
$\rho(K-1)2^{-\frac{B_{\text{f}}}{N_{t}-1}}$. On the other hand,
in the undesired small $B_{\text{c}}$ regimes, the second term in
(\ref{eq:interference-leakage-K-user-approx}) dominates, which represents
the residual interference due to poor quantization for \ac{csi} exchange
among users. 

\section{Adaptive CSI Exchange for \\Non-identical Channels}

\label{sec:CSI-exchange-design}


In this section, we study the case of non-identical channels, where
users may have different path loss $l_{k}$ and different channel
covariance matrices $\mathbf{R}_{k}$. In this scenario, it is not
efficient to distribute equal bits to the users for CSI exchange.
The intuitions are as follows. First, some users may be in the interference
limited region and require the other users to know their channels
for interference aware precoding, whereas some other users may be
in the noise limited region and inter-user interference is not an
essential issue for them. Second, when two users have non-overlapping
signal subspaces, they do not need to exchange the CSI, because there
is no interference for each other even under \ac{mrc} precoding.
Therefore, the users should have different CSI exchange strategies
according to the global CSI statistics $\{l_{k},\mathbf{R}_{k}\}$
and the whole D2D resources $B_{\text{tot}}$ bits should be smartly
partitioned over all the user pairs. 

We first specify the precoding strategy for cooperative precoder feedback
scheme. Then, we elaborate the proposed CSI exchange strategy and
analyze the interference leakage for the cooperative precoder feedback
scheme. Based on this, we derive the optimal bit partition for CSI
exchange.

\subsection{Precoding Strategy}

\label{subsec:individual-precoding}

\mysubsubnote{The codebook design is heuristic here. The intuition is that when users have non-overlapping signal subspaces, the proposed precoder feedback scheme degnerates to the CSI feedback scheme. Then such codebook is (statistically) optimal. Moreover, this sets a fair comparison with the CSI feedback scheme. The comparison is only on the feedback schem, but not the codebook design.}

Consider the following precoder codebook 
\begin{equation}
\mathcal{C}_{k}^{\text{w}}=\Big\{\mathbf{u}_{i}:\mathbf{u}_{i}=\mathbf{R}_{k}^{\frac{1}{2}}\bm{\xi}_{i}/\|\mathbf{R}_{k}^{\frac{1}{2}}\bm{\xi}_{i}\|_{2},\,i=1,2,\dots,2^{B_{\text{f}}}\Big\}\label{eq:precoder-codebook}
\end{equation}
where $\bm{\xi}_{i}$ are random vectors following complex Gaussian
distribution $\mathcal{CN}(\mathbf{0},\mathbf{I})$. 

After the CSI exchange, user $k$ chooses the precoder $\mathbf{w}_{k}^{\text{c}}$
from precoder codebook $\mathcal{C}_{k}^{\text{w}}$ to maximize the
\ac{slnr} as follows 
\begin{equation}
\mathbf{w}_{k}^{\text{c}}=\mathcal{W}_{k}(\hat{\mathbf{H}}_{k})\triangleq\arg\max_{\mathbf{w}\in\mathcal{C}_{k}^{\text{w}}}\;\frac{|\mathbf{h}_{k}^{\text{H}}\mathbf{w}|^{2}}{\alpha+\sum_{j\neq k}|\hat{\mathbf{h}}_{j}^{(k)\text{H}}\mathbf{w}|^{2}}\label{eq:precoding-slnr}
\end{equation}
where $\alpha=K/P$.

The motivation to use \ac{slnr} precoder is that the \ac{slnr} precoding
has been shown to achieve good performance in multiuser MIMO systems
from low to high \ac{snr} \cite{peel2005vector,sadek2007leakage,joham2005linear,PatArDou12}.
In addition, there is a strong relation between \ac{slnr} precoding
and \ac{mmse} precoding.
\begin{remrk}
[Connection between \ac{slnr} Precoding and \Ac{mmse} Precoding]\label{rem:connection}
Consider precoding strategies in the continuous domain (i.e., without
constrained in the precoder codebook $\mathcal{C}_{k}^{\text{w}}$).
The transmit precoding vector that satisfies \ac{mmse} criteria is
given by 
\begin{equation}
\mathbf{w}_{k}^{\mbox{\scriptsize MMSE}}=\sqrt{\Psi_{k}}\left(\hat{\mathbf{H}}_{k}\hat{\mathbf{H}}_{k}^{\text{H}}+\alpha\mathbf{I}\right)^{-1}\mathbf{h}_{k}\label{eq:w-mmse}
\end{equation}
where $\Psi_{k}$ is a normalizing factor such that $\|\widetilde{\mathbf{w}}_{k,\mbox{\scriptsize MMSE}}\|^{2}=1$.
On the other hand, the \ac{slnr} precoding vector is given by 
\begin{equation}
\mathbf{w}_{k}^{\mbox{\scriptsize SLNR}}=\arg\max_{\|\mathbf{w}\|^{2}=1}\frac{|\mathbf{h}_{k}^{\text{H}}\mathbf{w}|^{2}}{\sum_{j\neq k}|\hat{\mathbf{h}}_{j}^{(k)\text{H}}\mathbf{w}|^{2}+\alpha}.\label{eq:w-slnr}
\end{equation}
It was shown in \cite{PatArDou12} that the \ac{mmse} precoder in
(\ref{eq:w-mmse}) is equivalent to the \ac{slnr} precoder in (\ref{eq:w-slnr})
up to a complex scaling, i.e., $\mathbf{w}_{k}^{\mbox{\scriptsize MMSE}}=c_{k}\mathbf{w}_{k}^{\mbox{\scriptsize SLNR}}$.
\end{remrk}

\subsection{CSI Exchange Strategy}

\label{subsec:quantization-CSI-exchange}

The proposed CSI exchange strategy consists of two components, namely,
subspace projection for dimension reduction, and D2D quantizer for
bit partition among different user pairs. 

\subsubsection{Subspace Projection}

We propose a channel quantization method for CSI exchange based on
\emph{signal subspace projection}. The strategy consists of two steps.
\begin{itemize}
\item Subspace projection: To share the channel $\mathbf{h}_{k}$ to user
$j$, user $k$ first computes the partial channel 
\begin{equation}
\mathbf{g}_{k}^{(j)}=\frac{1}{\sqrt{l_{k}}}\mathbf{U}_{j}^{\text{H}}\mathbf{h}_{k}\label{eq:interf-subspace-projection-vector}
\end{equation}
where $\mathbf{U}_{j}$ is a $N_{t}\times\bar{M}_{j}$ matrix that
contains the $\bar{M}_{j}$ dominant eigenvectors of the covariance
matrix $\mathbf{R}_{j}$ of user $j$. 
\item Quantization: The partial channel $\mathbf{g}_{k}^{(j)}$ is quantized
into $\hat{\mathbf{g}}_{k}^{(j)}$ using $b_{kj}$ bits and transmitted
to user $j$.
\end{itemize}
User $j$ obtains the channel from user $k$ as $\hat{\mathbf{h}}_{k}^{(j)}=\sqrt{l_{k}}\mathbf{U}_{j}\hat{\mathbf{g}}_{k}^{(j)}$. 

\begin{remrk}
[Intuitive Interpretation] The intuition of the subspace projection
is that, only the portion of the channel that lies in the overlapping
signal subspace needs to be exchanged. To see this, rewrite the channel
of user $k$ as 
\begin{align*}
\mathbf{h}_{k} & =\mathbf{U}_{j}\mathbf{U}_{j}^{\text{H}}\mathbf{h}_{k}+(\mathbf{I}-\mathbf{U}_{j}\mathbf{U}_{j}^{\text{H}})\mathbf{h}_{k}\\
 & =\mathbf{h}_{k}^{(j)}+\mathbf{h}_{k}^{(j)\perp}
\end{align*}
where $\mathbf{h}_{k}^{(j)}$, which can be written as $\mathbf{h}_{k}^{(j)}=\sqrt{l_{k}}\mathbf{U}_{j}\mathbf{g}_{k}^{(j)}$,
is the portion of $\mathbf{h}_{k}$ that lies in the overlapping signal
subspace of users $k$ and $j$, whereas, $\mathbf{h}_{k}^{(j)\perp}$
is orthogonal to the overlapping signal subspace. From the construction
of precoder codebook $\mathcal{C}_{j}^{\text{w}}$ in (\ref{eq:precoder-codebook}),
the precoder $\mathbf{w}_{j}^{\text{c}}$ lies in the subspace spanned
by $\mathbf{U}_{j}$. As a result, $|(\mathbf{h}_{k}^{(j)\perp})^{\text{H}}\mathbf{w}_{j}^{\text{c}}|=0$
almost surely, and hence there is no need to transmit $\mathbf{h}_{k}^{(j)\perp}$
to user $j$.
\end{remrk}

\subsubsection{D2D Quantizer}

Note that in the conventional CSI feedback scheme, the CSI is used
for both signal enhancement and interference mitigation, whereas in
the proposed precoder feedback scheme (\ref{eq:precoding-slnr}),
the CSI exchanged among users is for interference mitigation only.
As a result, not all the users require the same level of CSI quality,
depending on the propagation scenarios such as signal subspace and
path loss. 

The concept of D2D quantizer for CSI exchange among users is highlighted
as follows.
\begin{definition}
[D2D Quantizer] A D2D quantizer $\mathcal{Q}(\{b_{kj}\})$ with
total bits $B_{\text{tot}}$ consists of bit partition $\{b_{kj}:\sum_{k=1}^{K}\sum_{j\neq k}b_{kj}=B_{\text{tot}}\}$
and a set of individual quantizers $\mathcal{Q}_{kj}$ with rate $b_{kj}$
that map the partial channel $\mathbf{g}_{k}^{(j)}$ to $\hat{\mathbf{g}}_{k}^{(j)}$. 
\end{definition}

There are many techniques to design the quantizers $\mathcal{Q}_{kj}$.
For example, for small number of bits $b_{kj}$, codebook based vector
quantization techniques can be used \cite{au2007performance,love2006limited,jindal2006mimo,yoo2007multi,bhagavatula2011adaptive,yuan2012bit}.
Here, we choose entropy-coded scalar quantization for elaboration
\cite{gersho2012vector}, because it is easier to scale to moderate
or large number of bits $b_{kj}$ for the scenario of CSI exchange
via D2D. 

\subsubsection{CSI Exchange using Entropy-coded Scalar Quantization}

The entropy-coded scalar quantizer works as follows.

First, \ac{klt} is applied to de-correlate the entries of the vector
$\mathbf{g}_{k}^{(j)}$. Let $\mathbf{G}_{kj}\triangleq\mathbb{E}\{\mathbf{g}_{k}^{(j)}\mathbf{g}_{k}^{(j)\text{H}}\}$
be the covariance matrix of the partial channel $\mathbf{g}_{k}^{(j)}$
and denote the eigen decomposition of $\mathbf{G}_{kj}$ as $\mathbf{G}_{kj}=\mathbf{U}_{kj}^{\text{H}}\bm{\Lambda}_{kj}\mathbf{U}_{kj}$,
where $\bm{\Lambda}_{kj}=\mbox{diag}(\lambda_{kj}^{(1)},\lambda_{kj}^{(2)},\dots,\lambda_{kj}^{(M_{kj})})$
is a diagonal matrix that contains the eigenvalues $\{\lambda_{kj}^{(i)}\}$
of $\mathbf{G}_{kj}$ in descending order. The \ac{klt} of $\mathbf{g}_{k}^{(j)}$
is given by
\begin{equation}
\mathbf{q}_{k}^{(j)}=\mathbf{U}_{kj}^{\text{H}}\mathbf{g}_{k}^{(j)}.\label{eq:klt}
\end{equation}
Note that there is a dimension reduction $M_{kj}<\bar{M}_{j}$ when
the subspaces of user $k$ and $j$ are only partially overlapped.

With the \ac{klt}, the $i$th element of $\mathbf{q}_{k}^{(j)}$
is $\mathcal{CN}(0,\lambda_{kj}^{(i)})$ distributed, and is uncorrelated
with the other elements of $\mathbf{q}_{k}^{(j)}$. Then, a scalar
quantizer is designed to quantize each element of $\mathbf{q}_{k}^{(j)}$,
and at the same time, lossless code (such as Hamming code) is applied
to encode the output of the quantizer, such that the average output
bit rate approaches to the entropy of the quantizer, where the entropy
is constrained to be $b_{kj}$. 

Finally, at user $j$, the channel of user $k$ is reconstructed as
$\hat{\mathbf{h}}_{k}^{(j)}=\sqrt{l_{k}}\mathbf{U}_{j}\hat{\mathbf{g}}_{k}^{(j)}=\sqrt{l_{k}}\mathbf{U}_{j}\mathbf{U}_{kj}\hat{\mathbf{q}}_{k}^{(j)}$.

Define the distortion of $\hat{\mathbf{g}}_{k}^{(j)}$ as the squared
error given by 
\begin{equation}
D_{kj}\triangleq\mathbb{E}\left\{ \|\mathbf{g}_{k}^{(j)}-\hat{\mathbf{g}}_{k}^{(j)}\|_{2}^{2}\right\} .\label{eq:distortion-g}
\end{equation}
The distortion-rate function $D_{kj}(b_{kj})$ is defined as the theoretical
minimum distortion $D_{kj}$ under $b_{kj}$ bits. As a direct application
of Shannon's distortion-rate theory \cite[Theorem 13.3.3]{cover2012elements},
the distortion-rate function for the above quantizer can be given
in the following lemma.
\begin{lemma}
[Distortion-rate] The distortion-rate function $D_{kj}(b_{kj})$
is given by 
\begin{equation}
D_{kj}(b_{kj})=m_{kj}^{*}\Bigg(\prod_{i=1}^{m_{kj}^{*}}\lambda_{kj}^{(i)}\Bigg)^{\frac{1}{m_{kj}^{*}}}2^{-\frac{b_{kj}}{m_{kj}^{*}}}+\sum_{i=m_{kj}^{*}+1}^{M_{kj}}\lambda_{kj}^{(i)}\label{eq:rate-distortion-function}
\end{equation}
where $m_{kj}^{*}\leq M_{kj}$ is a positive integer such that $\sum_{i=1}^{m_{kj}^{*}}r_{i}(m_{kj}^{*})=b_{kj}$,
in which $r_{i}(m)=\max\Big\{0,\frac{b_{kj}}{m}+\log_{2}\big[\lambda_{kj}^{(i)}/(\prod_{i=1}^{m}\lambda_{kj}^{(i)})^{\frac{1}{m}}\big]\Big\}$. 
\end{lemma}

\begin{remrk}
[Achievability] The distortion-rate function $D_{kj}(b_{kj})$ in
(\ref{eq:rate-distortion-function}) can be roughly achieved by applying
infinite level uniform scalar quantizer \cite{marco2006low} with
reverse water-filling bit allocation to distribute $b_{kj}$ bits
over the elements of $\mathbf{q}_{k}^{(j)}$ \cite[Theorem 13.3.3]{cover2012elements},
and at the same time, encoding the output of the quantizer using lossless
codes (such as Hamming code). Note that the operational distortion-rate
$\hat{D}_{kj}(b_{kj})$ under such method asymptotically approaches
to the Shannon's distortion-rate function (\ref{eq:rate-distortion-function})
in low resolution regime (small $b_{kj}$) \cite{marco2006low}. In
high resolution regime, it requires additional $0.25$ bit per real
dimension to achieve the same distortion as $D_{kj}(b_{kj})$. Nevertheless,
it is still insightful to apply $D_{kj}(b_{kj})$ in (\ref{eq:rate-distortion-function})
to analyze the performance and optimize $b_{kj}$ in the remaining
part of the paper. 
\end{remrk}

\subsection{Bit Partition for CSI Exchange}

\label{subsec:bit-partition}

Intuitively, the bit partition for CSI exchange should mainly depend
on the channel statistics $\{l_{k},\mathbf{R}_{k}\}$ and the quantizers
$\mathcal{Q}_{kj}$, but should not be quite affected by the number
of feedback bits $B_{\text{f}}$. To make the analysis tractable and
isolate the impact of $B_{\text{f}}$, the concept of \emph{virtual
SLNR} is introduced as follows.
\begin{definition}
[Operational SLNR] The operational SLNR of user $k$ is defined
as
\[
\gamma_{k}(\mathbf{H},\mathcal{Q},\mathcal{C}_{k}^{\text{w}})\triangleq\frac{|\mathbf{h}_{k}^{\text{H}}\mathbf{w}_{k}|^{2}}{\sum_{j\neq k}|\mathbf{h}_{j}^{\text{H}}\mathbf{w}_{k}|^{2}+\alpha}
\]
where $\alpha>0$ is some regularization parameter, and $\mathbf{w}_{k}=\mathbf{w}_{k}^{\text{c}}(\hat{\mathbf{H}}_{k},\mathcal{C}_{k}^{\text{w}})$
is the precoder from the cooperative feedback based on the partial
global CSI $\hat{\mathbf{H}}_{k}$ (depending on the CSI exchange
quantizer $\mathcal{Q}$) and the precoder codebook $\mathcal{C}_{k}^{\text{w}}(B_{\text{f}})$,
which contains $2^{B_{\text{f}}}$ precoding vectors. 
\end{definition}

\begin{definition}
[Virtual SLNR $\bar{\Gamma}_{k}$] Given the bit partition $\{b_{kj}\}$,
SLNR $\Gamma_{k}$ is achievable if there exists a D2D quantizer $\mathcal{Q}(\{b_{kj}\})$
and a sequence of precoder codebooks $\mathcal{C}_{k}^{\text{w}}(B_{\text{f}})$
such that $\lim_{B_{\text{f}}\to\infty}\mathbb{E}\{\gamma_{k}(\mathbf{H},\mathcal{Q},\mathcal{C}_{k}^{\text{w}})\}\geq\Gamma_{k}$.
The virtual SLNR $\bar{\Gamma}_{k}(\{b_{kj}\})$ is the supremum of
the achievable SLNR $\Gamma_{k}$.
\end{definition}

The virtual SLNR $\bar{\Gamma}_{k}(\{b_{kj}\})$ is a function to
characterize the theoretical performance of the bit partition $\{b_{kj}\}$
for CSI exchange. It isolates the impacts from the precoder codebook
$\mathcal{C}_{k}^{\text{w}}$ and the parameter $B_{\text{f}}$. Ideally,
the virtual SLNR $\bar{\Gamma}_{k}(\{b_{kj}\})$ can be achieved by
SLNR precoding in the continuous domain $\|\mathbf{w}_{k}\|=1$ (as
in (\ref{eq:w-slnr})) and optimal quantizers $\mathcal{Q}_{kj}$
for CSI exchange. As a result, the virtual SLNR $\bar{\Gamma}_{k}(\{b_{kj}\})$
serves as a good performance metric for bit partition. 

Specifically, the bit partition that maximizes the virtual SLNR is
formulated as follows 
\begin{align}
\underset{\{b_{kj}\geq0\}}{\mbox{maximize}} & \quad\sum_{k=1}^{K}\log(\bar{\Gamma}_{k}(\{b_{kj}\}))\label{eq:bit-allocation-problem}\\
\mbox{subject to} & \quad\sum_{k=1}^{K}\sum_{j\neq k}b_{kj}=B_{\text{tot}}\nonumber 
\end{align}
where the $\log$ function in the objective is to impose proportional
fairness among users. 

\subsubsection{Virtual SLNR Lower Bound}

The explicit expression of virtual SLNR in (\ref{eq:bit-allocation-problem})
is difficult to obtain. Instead, a lower bound can be derived as follows.

We first study the model of partial CSI \textbf{$\hat{\mathbf{g}}_{k}^{(j)}$}.
\begin{lemma}
[CSI Model under High Resolution CSI Exchange]\label{lem:partial-CSI-model}
For sufficiently large $b_{kj}$, the CSI $\mathbf{h}_{k}^{(j)}$
can be statistically written as 
\end{lemma}

\begin{equation}
\mathbf{h}_{k}^{(j)}=\beta_{kj}\mathbf{U}_{j}\hat{\mathbf{g}}_{k}^{(j)}+\tau_{kj}\mathbf{U}_{j}\mathbf{s}_{k}^{(j)}\label{eq:partial-CSI-model}
\end{equation}
where $\beta_{kj}=\sqrt{l_{k}}\hat{\mathbf{g}}_{k}^{(j)\text{H}}\mathbf{g}_{k}^{(j)}/\|\hat{\mathbf{g}}_{k}^{(j)}\|^{2}$,
$\mathbf{s}_{k}^{(j)}$ is a unit norm isotropic random vector that
is independent to $\tau_{kj}$ and orthogonal to $\hat{\mathbf{g}}_{k}^{(j)}$.
Moreover, 
\begin{equation}
\mathbb{E}\{\tau_{kj}^{2}\}\leq l_{k}M_{kj}\Bigg(\prod_{i=1}^{M_{kj}}\lambda_{kj}^{(i)}\Bigg)^{\frac{1}{M_{kj}}}2^{-\frac{b_{kj}}{M_{kj}}}.\label{eq:tau-distortion}
\end{equation}

\begin{IEEEproof}
Please refer to Appendix \ref{sec:app-proof-lem-partial-CSI-model}.
\end{IEEEproof}

Using the partial CSI model in Lemma \ref{lem:partial-CSI-model},
the lower bound of the virtual SLNR can be derived as follows.
\begin{lemma}
[Virtual SLNR Lower Bound]\label{lem:Virtual-SLNR-Lower-bound}
For sufficiently large $b_{kj}$, the virtual SLNR $\bar{\Gamma}_{k}$
is lower bounded by 
\begin{equation}
\bar{\Gamma}_{k}(\{b_{kj}\})\geq l_{k}\sum_{i=K}^{N_{t}}\lambda_{k}^{(i)}\bigg[\sum_{j\neq k}l_{j}\Big(\prod_{m=1}^{M_{jk}}\lambda_{jk}^{(m)}\Big)^{\frac{1}{M_{jk}}}2^{-\frac{b_{jk}}{M_{jk}}}+\alpha\bigg]^{-1}.\label{eq:virtual-slnr-lower-bound}
\end{equation}
\end{lemma}

\begin{IEEEproof}
Please refer to Appendix \ref{sec:app-proof-lem-virtual-slnr-lower-bound}.
\end{IEEEproof}

%

%


\subsubsection{Optimal Bit Partition}

With the explicit expression on the virtual SLNR lower bound, the
bit partition problem via SLNR maximization (\ref{eq:bit-allocation-problem})
can be reformulated as maximizing the virtual SLNR lower bound (\ref{eq:virtual-slnr-lower-bound}). 

Note that since 
\begin{align}
 & \log\Bigg(l_{k}\sum_{i=K}^{N_{t}}\lambda_{k}^{(i)}\bigg(\sum_{j\neq k}\omega_{jk}2^{-\frac{b_{jk}}{M_{jk}}}+\alpha\bigg)^{-1}\Bigg)\label{eq:log-big-one}\\
 & \quad=\log\bigg(l_{k}\sum_{i=K}^{N_{t}}\lambda_{k}^{(i)}\bigg)-\log\bigg(\sum_{j\neq k}\omega_{jk}2^{-\frac{b_{jk}}{M_{jk}}}+\alpha\bigg)\label{eq:log-two-terms}
\end{align}
where $\omega_{jk}\triangleq l_{j}\bigg(\prod_{i=1}^{M_{jk}}\lambda_{jk}^{(i)}\bigg)^{\frac{1}{M_{jk}}}$,
maximizing (\ref{eq:log-big-one}) over $\{b_{kj}\}$ is equivalent
to minimizing the second term of (\ref{eq:log-two-terms}). Specifically,
the bit partition problem can be reformulated as follows 
\begin{align}
\underset{\{b_{kj}\geq0\}}{\mbox{minimize}} & \quad\sum_{k=1}^{K}\log\Bigg(\sum_{j\neq k}\omega_{jk}2^{-\frac{b_{jk}}{M_{jk}}}+\alpha\Bigg)\label{eq:bit-partition-problem-minimization}\\
\mbox{subject to} & \quad\sum_{k=1}^{K}\sum_{j\neq k}b_{kj}=B_{\text{tot}}.\nonumber 
\end{align}

The minimization problem (\ref{eq:bit-partition-problem-minimization})
is convex and the optimal solution can be obtained. 

Let $\mathbf{x}_{k}=(x_{1k},x_{2k},\dots,x_{k-1,k},x_{k+1,k},\dots,x_{Kk})^{\text{T}}$
and $\bm{\omega}_{k}=(\omega_{1k},\omega_{2k},\dots,\omega_{k-1,k},\omega_{k+1,k},\dots,\omega_{Kk})^{\text{T}}$
be two vectors each with $K-1$ entries. 

\begin{thm}
[Optimal Bit Partition]\label{thm:Optimal-Bit-Partition} The optimal
bit partition that minimizes the virtual SLNR lower bound in (\ref{eq:bit-partition-problem-minimization})
is given by 
\begin{equation}
b_{jk}=\left[-M_{jk}\log_{2}x_{jk}\right]^{+}\label{eq:optimal-bit-partition-solution}
\end{equation}
where $[x]^{+}=\max\{0,x\}$, and $x_{jk}$ is an entry in vector
$\mathbf{x}_{k}$ given by 
\begin{equation}
\mathbf{x}_{k}=\mu\alpha(\mathbf{A}_{k}-\mu\mathbf{1}\bm{\omega}_{k}^{\text{T}})^{-1}\mathbf{1}\label{eq:optimal-bit-partition-solution-auxilliary}
\end{equation}
in which, 
\[
\mathbf{A}_{k}=\ln2\cdot\mbox{diag}(M_{1k}^{-1},M_{2k}^{-1},\dots,M_{k-1,k}^{-1},M_{k+1,k}^{-1},\dots,M_{Kk}^{-1})
\]
the parameter $\mu$ is a non-negative variable chosen such that $\sum_{k=1}^{K}\sum_{j\neq k}b_{kj}(\mu)=B_{\text{tot}}$,
and $\mathbf{1}$ is a $K-1$ dimensional column vector with all the
entries being $1$'s, 
\end{thm}

\begin{IEEEproof}
Please refer to Appendix \ref{sec:app-proof-thm-optimal-bit-partition}.
\end{IEEEproof}

Note that, since the problem is convex, the parameter $\mu$ can be
found using bisection search, which converges very fast. 

The results in Theorem \ref{thm:Optimal-Bit-Partition} suggests that
the optimal bit partition for \ac{csi} exchange varies according
to the path loss $l_{k}$, the dimension $M_{jk}$ of the interference
subspace between user $k$ and $j$, as well as the eigenvalues of
the covariance $\widetilde{\mathbf{R}}_{jk}$ of the overlapping subspace.

\section{Numerical Results}

\label{sec:numerical}

In this section, we evaluate the performance of the precoding feedback
scheme with adaptive CSI exchange when users have different CSI statistics. 

Consider a single cell downlink massive MIMO system with $N_{t}=60$
antennas at the \ac{bs} serving $K=2$ single antenna users. The
noise variance is normalized to $1$. The one-ring model \cite{Forenza:2007cr,Zhang:2007zr}
on \ac{ula} is used for the channel modeling. The angular spread
is $15$ degrees and the power angular spectrum density follows a
truncated Gaussian distribution centered at the mean azimuth direction
of the user.\footnote{The numerical results for the identically uncorrelated channels can
be found in \cite{chen2015precoder}.} Each user has $B_{\text{f}}=6$ bits to feedback the precoder or
the CSI to the \ac{bs}, and the two users have in total $B_{\text{tot}}=80$
bits for CSI exchange in the precoder feedback schemes. 

The following CSI exchange, feedback, and precoding schemes are evaluated
\begin{itemize}
\item \textbf{Baseline 1 (CSI Feedback):} \Ac{mmse} precoding is computed
by the \ac{bs} according to the CSI feedback from each user in $B_{\text{f}}$
bits.
\item \textbf{Baseline 2 (Precoder feedback with naive CSI exchange):} The
\ac{csi} is quantized and exchanged according to each user's own
\ac{csi} statistics, using $B_{\text{tot}}/2$ bits for each user.
The precoder is computed according to (\ref{eq:precoding-slnr}) and
fed back to the BS.
\item \textbf{Proposed (Precoder feedback with adaptive CSI exchange): }The
\ac{csi} is quantized and exchanged according to the proposed strategy
in Section \ref{subsec:quantization-CSI-exchange} with adaptive bit
partition for each user as in Section \ref{subsec:bit-partition}.
The precoder is computed according to (\ref{eq:precoding-slnr}) and
fed back to the BS.
\end{itemize}

\subsection{Heterogeneous Path Loss}

Consider the two users are near to each other and therefore they share
the same signal subspace. However user 2 suffers from larger path
loss due to additional blockage.\footnote{For example, user $1$ is outdoor and user $2$ is indoor. }
As a result, the two user have the same signal subspace, but user
$2$ suffers from larger path loss. 

\begin{figure}
\begin{centering}
\psfragscanon
\psfrag{5}[][][0.7]{5}
\psfrag{6}[][][0.7]{6}
\psfrag{8}[][][0.7]{8}
\psfrag{7}[][][0.7]{7}
\psfrag{9}[][][0.7]{9}
\psfrag{10}[][][0.7]{10}
\psfrag{12}[][][0.7]{12}
\psfrag{14}[][][0.7]{14}
\psfrag{5}[][][0.7]{5}
\psfrag{10}[][][0.7]{10}
\psfrag{15}[][][0.7]{15}
\psfrag{20}[][][0.7]{20}
\psfrag{25}[][][0.7]{25}
\psfrag{30}[][][0.7]{30}
\psfrag{0}[][][0.7]{0}
\psfrag{Adaptive CSI Exch}[Bl][Bl][0.7]{Proposed}
\psfrag{BS-MMSE}[Bl][Bl][0.7]{CSI Feedback}
\psfrag{D2D Naive}[Bl][Bl][0.7]{Naive CSI Exch}
\psfrag{Sum Rate (bps/Hz)}[][][0.8]{Sum rate (bps/Hz)}
\psfrag{Blockage of the 2nd user (dB)}[][][0.8]{Additional blockage of user 2 (dB)}\includegraphics{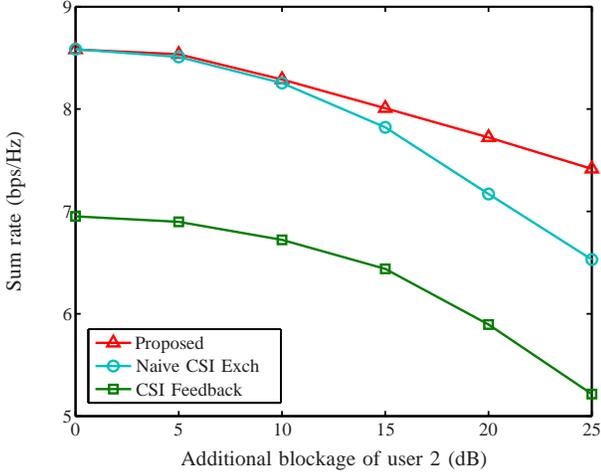}
\par\end{centering}
\caption{\label{fig:PL_sum_rate} Sum rate versus additional blockage of user
2 under total transmission power $P=20$ dB.}
\end{figure}

Fig. \ref{fig:PL_sum_rate} shows the sum rate versus additional blockage
of user 2 under total transmission power $P=20$ dB. Specifically,
the path loss of user $1$ is normalized to $1$, and the path loss
of user 2 is equal to the blockage. First, both precoder feedback
schemes significantly outperform the CSI feedback scheme. Second,
the proposed precoder feedback with adaptive CSI exchange outperforms
the naive CSI exchange scheme. This is because, user $2$ is in the
noise limited region, and hence it is not necessary for user $2$
to inform its CSI $\mathbf{h}_{2}$ to user $1$ for interference
mitigation. On the other hand, user $1$ wishes user $2$ to know
its CSI $\mathbf{h}_{1}$ for interference aware precoding, since
user $1$ is in interference limited region. Therefore, equal bit
partition in the naive CSI exchange scheme is not efficient. The bit
partition results for the proposed adaptive CSI exchange scheme is
summarized in Table \ref{tab:bit-partition}.

\begin{table}
\renewcommand{\arraystretch}{1.6}
\begin{centering}
\begin{tabular}{|c|c|c|c|c|c|c|}
\hline 
Blockage of user 2 (dB) & 0 & 5 & 10 & 15 & 20 & 25\tabularnewline
\hline 
\hline 
Bits for user 1 to quantize $\mathbf{h}_{1}^{(2)}$ & 40 & 49 & 58 & 67 & 76 & 80\tabularnewline
Bits for user 2 to quantize $\mathbf{h}_{2}^{(1)}$ & 40 & 31 & 22 & 13 & 4 & 0\tabularnewline
\hline 
\end{tabular}
\par\end{centering}
\caption{\label{tab:bit-partition} Bit partition for adaptive CSI exchange
according to additional blockage of user 2. }
\end{table}

\subsection{Heterogeneous Signal Subspace}

Consider that the two users have the same path loss (normalized to
$1$), but the users are separated by $10$ meters and away from the
\ac{bs} by 60 meters. As a result, they have different signal subspace
due to the limited angular spread. 

\subsubsection{Sum Rate Performance}

\begin{figure}
\begin{centering}
\psfragscanon
\psfrag{4}[][][0.7]{4}
\psfrag{6}[][][0.7]{6}
\psfrag{8}[][][0.7]{8}
\psfrag{10}[][][0.7]{10}
\psfrag{12}[][][0.7]{12}
\psfrag{14}[][][0.7]{14}
\psfrag{5}[][][0.7]{5}
\psfrag{10}[][][0.7]{10}
\psfrag{15}[][][0.7]{15}
\psfrag{20}[][][0.7]{20}
\psfrag{25}[][][0.7]{25}
\psfrag{30}[][][0.7]{30}
\psfrag{0}[][][0.7]{0}
\psfrag{Adaptive CSI Exch}[Bl][Bl][0.7]{Proposed}
\psfrag{BS-MMSE}[Bl][Bl][0.7]{Naive CSI Exch}
\psfrag{D2D Naive}[Bl][Bl][0.7]{CSI Feedback}
\psfrag{Sum Rate (bps/Hz)}[][][0.8]{Sum rate (bps/Hz)}
\psfrag{Total TX Power (dB)}[][][0.8]{Total TX power $P$ (dB)}\includegraphics{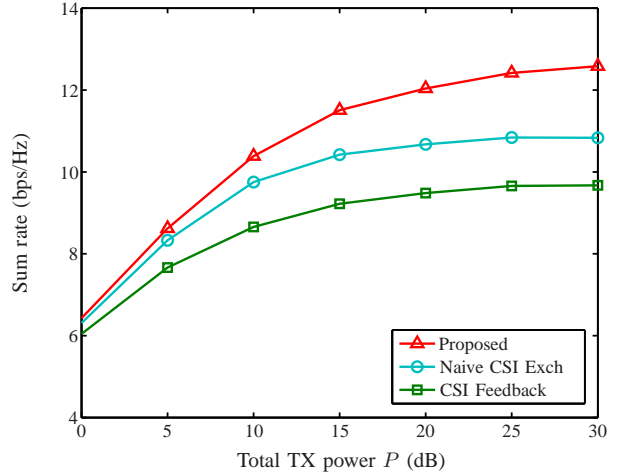}
\par\end{centering}
\caption{\label{fig:rate-snr} Sum rate versus the total transmission power. }
\end{figure}

Fig. \ref{fig:rate-snr} shows the sum rate versus the total transmission
power. First, both precoder feedback schemes outperform the CSI feedback
scheme. Second, the proposed precoder feedback with adaptive CSI exchange
outperforms the naive CSI exchange scheme, because the proposed scheme
quantizes the CSI using the statistics of both users. Specifically,
it only quantizes the portion of CSI that lies in the overlapping
signal subspace of the two users, and hence the quantization is more
efficient. 

\subsubsection{Feedback Saving}

Fig. \ref{fig:rate-fb} (a) demonstrates the sum rate versus the number
of bits $B_{\text{f}}$ per user for the feedback to the \ac{bs}
under total transmission power $P=10$ dB. Both precoder feedback
schemes outperform the CSI feedback scheme under $B_{\text{f}}=4$
to $12$ feedback bits. In particular, the proposed scheme saves almost
half of the bits for the feedback to the \ac{bs} under similar sum
rate performance as the CSI feedback scheme. 

\subsubsection{D2D Signaling Saving}

\begin{figure}
\begin{centering}
\subfigure[]{
\psfragscanon
\psfrag{4}[][][0.7]{4}
\psfrag{5}[][][0.7]{5}
\psfrag{6}[][][0.7]{6}
\psfrag{7}[][][0.7]{7}
\psfrag{8}[][][0.7]{8}
\psfrag{9}[][][0.7]{9}
\psfrag{10}[][][0.7]{10}
\psfrag{11}[][][0.7]{11}
\psfrag{12}[][][0.7]{12}
\psfrag{13}[][][0.7]{13}
\psfrag{Adaptive CSI Exch}[Bl][Bl][0.7]{Proposed}
\psfrag{D2D (naive)}[Bl][Bl][0.7]{Naive CSI Exch}
\psfrag{BS, MMSE}[Bl][Bl][0.7]{CSI Feedback}
\psfrag{Bits per UE to feedback to BS}[][][0.8]{Bits per user to feedback $B_{\mbox{\scriptsize f}}$}
\psfrag{Sum Rate (bps/Hz)}[][][0.8]{Sum rate (bps/Hz)}\includegraphics[width=1\columnwidth]{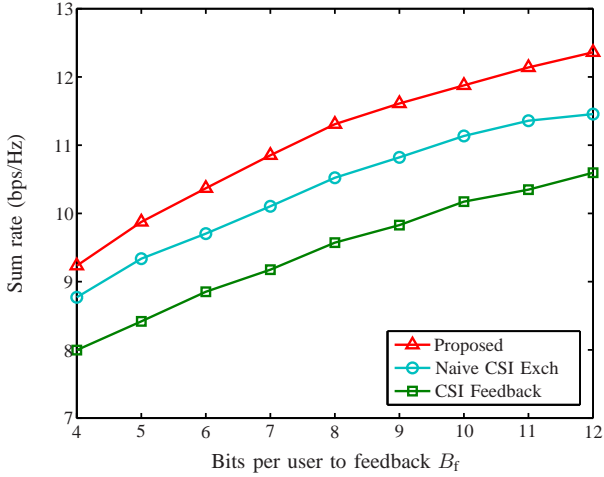}}
\par\end{centering}
\begin{centering}
\subfigure[]{
\psfragscanon
\psfrag{7}[][][0.7]{7}
\psfrag{8}[][][0.7]{8}
\psfrag{9}[][][0.7]{9}
\psfrag{10}[][][0.7]{10}
\psfrag{11}[][][0.7]{11}
\psfrag{20}[][][0.7]{20}
\psfrag{40}[][][0.7]{40}
\psfrag{60}[][][0.7]{60}
\psfrag{80}[][][0.7]{80}
\psfrag{100}[][][0.7]{100}
\psfrag{120}[][][0.7]{120}
\psfrag{BS-MMSE}[Bl][Bl][0.7]{CSI Feedback}
\psfrag{Naive CSI Exch}[Bl][Bl][0.7]{Naive CSI Exch}
\psfrag{Adaptive CSI Exch}[Bl][Bl][0.7]{Proposed}
\psfrag{Total TX Power (dB)}[][][0.8]{Total bits $B_{\mbox{\scriptsize tot}}$ for CSI exchange}
\psfrag{Sum Rate (bps/Hz)}[][][0.8]{Sum rate (bps/Hz)}\includegraphics[width=1\columnwidth]{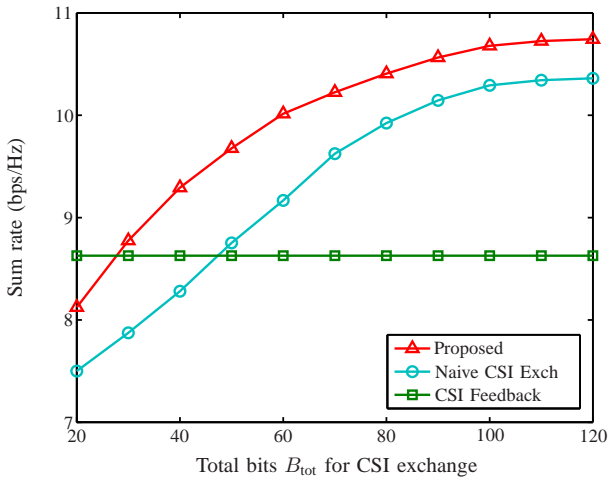}}
\par\end{centering}
\caption{\label{fig:rate-fb} Sum rate versus (a) the number of bits per user
$B_{\text{f}}$ for the feedback to the \ac{bs} and (b) total number
of bits $B_{\text{d}}$ for \ac{csi} exchange.}
\end{figure}

Fig. \ref{fig:rate-fb} (b) shows the sum rate versus total number
of bits $B_{\text{tot}}$ for \ac{csi} exchange under total transmission
power $P=10$ dB. The CSI feedback scheme is not affected by $B_{\text{tot}}$.
The result demonstrates that when there are sufficient number of bits
for \ac{csi} exchange, precoder feedback is preferred over CSI feedback.
Under limited feedback to the \ac{bs} and limited D2D signaling,
the proposed scheme saves one third to almost half of the bits for
\ac{csi} exchange as compared to the naive CSI exchange scheme. 

\section{Conclusions}

\label{sec:conclusion}

We proposed a cooperative precoder feedback strategy for multiuser
downlink transmission in \ac{fdd} massive MIMO systems. The strategy
consists of two phases. First, the users exploit reliable \ac{d2d}
communication to exchange the CSI, and second, the users individually
compute the precoder and feed back the precoder to the \ac{bs}. We
analyzed the interference leakage when users have identically uncorrelated
channel statistics. Our results showed that the precoder feedback
scheme can reduce the interference leakage to $1/(K-1)$ of the CSI
feedback scheme with \ac{zf} precoding. When users have non-identical
channel statistics, we developed novel adaptive CSI exchange strategy,
which exploits the global CSI statistics of the users. Optimal bit
partition algorithm was derived for CSI exchange in terms of maximizing
the virtual SLNR. Numerical results demonstrated that the proposed
precoder feedback scheme with adaptive CSI exchange significantly
outperforms the CSI feedback scheme in terms of higher throughput
and lower feedback. The results also showed that the proposed scheme
significantly saves the D2D overhead for CSI exchange.

\appendices

\section{Proof of Lemma \ref{lem:characterization-interference-leakage}}

\label{sec:app-proof-lem-IL-d2d-precoding}

We first note that, vectors $\mathbf{s}_{j}$ follow an isotropic
distribution in the $(N_{t}-1)$-dimensional subspace, because both
of the quantization vectors $\hat{\mathbf{g}}_{j}$ from the \ac{rvq}
codebook and the channel direction vectors $\mathbf{g}_{j}$ are isotropically
distributed in the $N_{t}$-dimensional sphere. Thus, for any unit
norm vector $\mathbf{w}$ independent of $\mathbf{s}_{j}$, $\mathbb{E}_{\mathcal{H},\mathcal{C}}\{\mathbf{s}_{j}^{\text{H}}\mathbf{w}\,\big|\mathbf{w}\}=0$.
Therefore, the following holds from the property of iterative expectation
\begin{align}
 & \mathbb{E}_{\mathcal{H},\mathcal{C}}\left\{ \mathbf{w}_{1}^{\text{H}}\hat{\mathbf{g}}_{j}\mathbf{s}_{j}^{\text{H}}\mathbf{w}_{1}\right\} \nonumber \\
 & \qquad\qquad=\mathbb{E}_{\mathcal{H},\mathcal{C}}\bigg\{\mathbf{w}_{1}^{\text{H}}\hat{\mathbf{g}}_{j}\,\mathbb{E}_{\mathcal{H},\mathcal{C}}\left\{ \mathbf{s}_{j}^{\text{H}}\mathbf{w}_{1}\bigg|\mathbf{w}_{1},\hat{\mathbf{g}}_{j}\right\} \bigg\}\nonumber \\
 & \qquad\qquad=0.\label{eq:zero-expectation-cross-term}
\end{align}
Similarly, $\mathbb{E}_{\mathcal{H},\mathcal{C}}\left\{ \mathbf{w}_{1}^{\text{H}}\mathbf{s}_{j}\hat{\mathbf{g}}_{j}^{\text{H}}\mathbf{w}_{1}\right\} =0$.
As $\mathbf{g}_{j}=\sqrt{1-Z_{j}}\hat{\mathbf{g}}_{j}+\sqrt{Z_{j}}\mathbf{s}_{j}$
from the definition of $Z_{j}$ and $\mathbf{s}_{j}$, the following
holds 
\begin{align*}
 & \mathbb{E}_{\mathcal{H},\mathcal{C}}\left\{ I_{1}\right\} \\
 & \;=\rho\sum_{j\neq1}\mathbb{E}_{\mathcal{H},\mathcal{C}}\left\{ \|\mathbf{h}_{j}\|^{2}|\mathbf{g}_{j}^{\text{H}}\mathbf{w}_{1}|^{2}\right\} \\
 & \;\stackrel{(a)}{=}\rho N_{t}\sum_{j\neq1}\mathbb{E}_{\mathcal{H},\mathcal{C}}\left\{ \bigg|\sqrt{1-Z_{j}}\hat{\mathbf{g}}_{j}^{\text{H}}\mathbf{w}_{1}+\sqrt{Z_{j}}\mathbf{s}_{j}^{\text{H}}\mathbf{w}_{1}\bigg|^{2}\right\} \\
 & \;=\rho N_{t}\sum_{j\neq1}\mathbb{E}_{\mathcal{H},\mathcal{C}}\bigg\{(1-Z_{j})\big|\hat{\mathbf{g}}_{j}^{\text{H}}\mathbf{w}_{1}\big|^{2}+Z_{j}\big|\mathbf{s}_{j}^{\text{H}}\mathbf{w}_{1}\big|^{2}\\
 & \qquad\quad+\sqrt{(1-Z_{j})Z_{j}}\bigg[\mathbf{w}_{1}^{\text{H}}\hat{\mathbf{g}}_{j}\mathbf{s}_{j}^{\text{H}}\mathbf{w}_{1}+\mathbf{w}_{1}^{\text{H}}\mathbf{s}_{j}\hat{\mathbf{g}}_{j}^{\text{H}}\mathbf{w}_{1}\bigg]\bigg\}\\
 & \;\stackrel{(b)}{=}\rho N_{t}\sum_{j\neq1}\mathbb{E}_{\mathcal{H},\mathcal{C}}\left\{ (1-Z_{j})\big|\hat{\mathbf{g}}_{j}^{\text{H}}\mathbf{w}_{1}\big|^{2}+Z_{j}\big|\mathbf{s}_{j}^{\text{H}}\mathbf{w}_{1}\big|^{2}\right\} .
\end{align*}
where $\stackrel{(a)}{=}$ is due to the fact that the channel magnitude
$\|\mathbf{h}_{j}\|^{2}$ is independent of both $\mathbf{g}_{j}$
(channel direction) and $\mathbf{w}_{1}$ (precoding based on $\{\mathbf{g}_{j}\}$),
and $\stackrel{(b)}{=}$ is due to (\ref{eq:zero-expectation-cross-term}).

\section{Proof of Theorem \ref{thm:interference-leakage-two-user}}

\label{sec:app-proof-thm-intereference-two-user}
\begin{lemma}
[Interference due to Discrete Precoding]\label{lem:IL-D2D-fb} The
random variable $\big|\hat{\mathbf{g}}_{2}^{\text{H}}\mathbf{w}_{1}^{\text{c}}\big|^{2}$
follows a beta distribution $\mathcal{B}(1,(N_{t}-1)2^{B_{\text{f}}})$
and its mean is given by $\left(1+(N_{t}-1)2^{B_{\text{f}}}\right)^{-1}$.
\end{lemma}

\begin{IEEEproof}
By the construction of a \ac{rvq} codebook $\mathcal{C}_{1}^{\text{w}}$,
the codewords $\mathbf{w}\in\mathcal{C}_{1}^{\text{w}}$ follow isotropic
distribution in the $N_{t}$-dimensional subspace. Thus $\big|\hat{\mathbf{g}}_{2}^{\text{H}}\mathbf{w}^{\text{c}}\big|^{2}$
follows $\mathcal{B}(1,N_{t}-1)$ distribution, with \ac{cdf} given
by 
\[
\mathbb{P}\left\{ \big|\hat{\mathbf{g}}_{2}^{\text{H}}\mathbf{w}^{\text{c}}\big|^{2}\leq x\right\} =1-(1-x)^{N_{t}-1}.
\]
As a result, 
\begin{align*}
\mathbb{P}\left\{ \big|\hat{\mathbf{g}}_{2}^{\text{H}}\mathbf{w}_{1}^{\text{c}}\big|^{2}\leq x\right\}  & =\mathbb{P}\left\{ \min_{\mathbf{w}^{\text{c}}\in\mathcal{C}_{1}^{\text{w}}}\,\big|\hat{\mathbf{g}}_{2}^{\text{H}}\mathbf{w}^{\text{c}}\big|^{2}\leq x\right\} \\
 & =1-\mathbb{P}\left\{ \big|\hat{\mathbf{g}}_{2}^{\text{H}}\mathbf{w}_{i}^{\text{c}}\big|^{2}>x,\,\forall\mathbf{w}_{i}^{\text{c}}\in\mathcal{C}_{1}^{\text{w}}\right\} \\
 & =1-\mathbb{P}\left\{ \big|\hat{\mathbf{g}}_{2}^{\text{H}}\mathbf{w}^{\text{c}}\big|^{2}>x\right\} ^{N_{\text{f}}}\\
 & =1-(1-x)^{(N_{t}-1)N_{\text{f}}}
\end{align*}
which means that $\min_{\mathbf{w}\in\mathcal{C}_{1}^{\text{w}}}\,\big|\hat{\mathbf{g}}_{2}^{\text{H}}\mathbf{w}\big|^{2}$
follows the beta distribution $\mathcal{B}(1,(N_{t}-1)N_{\text{f}})$,
where $N_{\text{f}}=|\mathcal{C}_{1}^{\text{w}}|=2^{B_{\text{f}}}$.

Moreover, the mean of a beta random variable $\mathcal{B}(\alpha,\beta)$
is given by $\alpha/(\alpha+\beta)$, and hence $\mathbb{E}\left\{ \big|\hat{\mathbf{g}}_{2}^{\text{H}}\mathbf{w}_{1}^{\text{c}}\big|^{2}\right\} =1/[1+(N_{t}-1)2^{B_{\text{f}}}]$.
\end{IEEEproof}

From Lemma \ref{lem:characterization-interference-leakage} and \ref{lem:IL-D2D-fb}
and the channel quantization error bounds in (\ref{eq:Zj-chanenl-quantization-error-bounds}),
we have 
\begin{align*}
 & \mathbb{E}_{\mathcal{H},\mathcal{C}}\left\{ I_{1}^{\text{c}}\right\} \\
 & \quad=\rho N_{t}\mathbb{E}_{\mathcal{H},\mathcal{C}}\left\{ (1-Z_{2})\big|\hat{\mathbf{g}}_{2}^{\text{H}}\mathbf{w}_{1}^{\text{c}}\big|^{2}+Z_{2}\big|\mathbf{s}_{2}^{\text{H}}\mathbf{w}_{1}^{\text{c}}\big|^{2}\right\} \\
 & \quad\leq\rho N_{t}\bigg[\bigg(1-\frac{N_{t}-1}{N_{t}}2^{-\frac{B_{\text{c}}}{N_{t}-1}}\bigg)\frac{1}{1+(N_{t}-1)2^{B_{\text{f}}}}\\
 & \qquad\qquad\qquad\qquad\qquad\qquad\quad+2^{-\frac{B_{\text{c}}}{N_{t}-1}}\times\frac{1}{N_{t}-1}\bigg]\\
 & \quad\leq\rho N_{t}\bigg[\bigg(1-\frac{N_{t}-1}{N_{t}}2^{-\frac{B_{\text{c}}}{N_{t}-1}}\bigg)\frac{2^{-B_{\text{f}}}}{N_{t}-1}+\frac{2^{-B_{\text{c}}/(N_{t}-1)}}{N_{t}-1}\bigg]\\
 & \quad=\frac{\rho N_{t}}{N_{t}-1}\left[2^{-B_{\text{f}}}+\bigg(1-\frac{N_{t}-1}{N_{t}}2^{-B_{\text{f}}}\bigg)2^{-\frac{B_{\text{c}}}{N_{t}-1}}\right].
\end{align*}

\section{Proof of Lemma \ref{lem:Asymptotic-chi-square-approx}}

\label{sec:app-proof-lem-asymptotic-chi-square-approx}

We first show that $Y_{j,i}\triangleq|\hat{\mathbf{h}}_{j}^{\text{H}}\mathbf{w}_{i}|^{2}\xrightarrow{\text{d}}\frac{1}{2}\chi^{2}(2)$,
as $N_{t}\to\infty$, where $\xrightarrow{\text{d}}$ denotes convergence
in distribution. Note that 
\[
Y_{j,i}=\|\hat{\mathbf{h}}_{j}\|^{2}|\hat{\mathbf{u}}_{j}^{\text{H}}\mathbf{w}_{i}|^{2}=\frac{1}{N_{t}}\|\hat{\mathbf{h}}_{j}\|^{2}\cdot N_{t}Z
\]
where $\hat{\mathbf{u}}_{j}=\hat{\mathbf{h}}_{j}/\|\hat{\mathbf{h}}_{j}\|$
and $Z=|\hat{\mathbf{u}}_{j}^{\text{H}}\mathbf{w}_{i}|^{2}$ is well-known
to follow the beta distribution $\mathcal{B}(1,N_{t}-1)$, because
$\hat{\mathbf{u}}_{j}$ is a unit norm $N_{t}$-dimensional vector
and $\mathbf{w}_{i}$ is random, isotropic, and independent to $\hat{\mathbf{u}}_{j}$.

Note that each element of $\mathbf{h}_{j}$ is \ac{iid} Gaussian
$\mathcal{C}\mathcal{N}(0,1)$. Then, by the Strong Law of Large Numbers,
\begin{align}
\frac{1}{N_{t}}\|\hat{\mathbf{h}}_{j}\|^{2} & =\frac{1}{N_{t}}\|\mathbf{h}_{j}\|^{2}\nonumber \\
 & =\frac{1}{N_{t}}\left(|h_{j1}|^{2}+|h_{j2}|^{2}+\cdots+|h_{jN_{t}}|^{2}\right)\xrightarrow{\text{a.s.}}1\label{eq:pf-lem-chi-square-approx-SLLN}
\end{align}
as $N_{t}\to\infty$, where $\xrightarrow{\text{a.s.}}$ denotes \ac{a.s.}
convergence.

Let $Z$ be a beta random variable following $\mathcal{B}(1,N_{t}-1)$
distribution. Then 
\begin{align*}
\lim_{N_{t}\to\infty}\mathbb{P}\left\{ N_{t}Z\leq x\right\}  & =\lim_{N_{t}\to\infty}\mathbb{P}\left\{ Z\leq\frac{x}{N_{t}}\right\} \\
 & =\lim_{N_{t}\to\infty}1-(1-\frac{x}{N_{t}})^{N_{t}-1}\\
 & =\lim_{N_{t}\to\infty}1-(1-\frac{x}{N_{t}})^{N_{t}}\frac{1}{1-x/N_{t}}\\
 & =1-e^{-x}
\end{align*}
On the other hand, $\mathbb{P}\left\{ \frac{1}{2}\chi^{2}(2)\leq x\right\} =1-e^{-x}$,
which shows that $N_{t}Z\xrightarrow{\text{d}}\frac{1}{2}\chi^{2}(2)$.
Using (\ref{eq:pf-lem-chi-square-approx-SLLN}), we can conclude that
$Y_{j}\xrightarrow{\text{d}}\frac{1}{2}\chi^{2}(2)$.

%
%

We then show that $Y_{j,i}$'s are mutually independent \ac{wrt}
$j$. First, from the independency of $\mathbf{h}_{j}$, the quantized
vectors $\hat{\mathbf{h}}_{j}$ are independent. In addition, from
$Y_{j,i}=\|\hat{\mathbf{h}}_{j}\|^{2}|\hat{\mathbf{u}}_{j}^{\text{H}}\mathbf{w}_{i}|^{2}$,
the random variables $|\hat{\mathbf{u}}_{j}^{\text{H}}\mathbf{w}_{i}|^{2}$
and $|\hat{\mathbf{u}}_{k}^{\text{H}}\mathbf{w}_{i}|^{2}$, $k\neq j$,
are mutually independent, because $\hat{\mathbf{u}}_{j}$ are independently
and isotropically distributed. These conclude that $Y_{j,i}$'s are
mutually independent \ac{wrt} $j$.

As a result, $\widetilde{Y}_{i}=\sum_{j\neq1}Y_{j,i}$ converges to
the sum of $K-1$ \ac{iid} $\frac{1}{2}\chi^{2}(2)$ random variables,
which is $\frac{1}{2}\chi^{2}(2(K-1))$. 

In addition, given $\{\hat{\mathbf{h}}_{j}\}_{j\neq1}$, $\widetilde{Y}_{i}$
and $\widetilde{Y}_{l}$ are independent. The independence of $|\hat{\mathbf{u}}_{j}^{\text{H}}\mathbf{w}_{i}|^{2}$
and $|\hat{\mathbf{u}}_{j}^{\text{H}}\mathbf{w}_{l}|^{2}$ follows
from the independence between isotropic random vectors $\mathbf{w}_{i}$.
As $\frac{1}{N_{t}}\|\hat{\mathbf{h}}_{j}\|^{2}\xrightarrow{\text{a.s.}}1$,
$\widetilde{Y}_{i}$ and $\widetilde{Y}_{l}$ become asymptotically
independent for large $N_{t}$. Hence, $(\widetilde{Y}_{1},\widetilde{Y}_{2},\dots,\widetilde{Y}_{N})$
converges to $\frac{1}{2}(X_{1},X_{2},\dots,X_{N})$ in distribution.

\section{Proof of Lemma \ref{lem:asymptotic-distribution}}

\label{sec:app-proof-lem-asymp-dist}

From Lemma \ref{lem:Asymptotic-chi-square-approx}, as $N_{t}\to\infty$,
$\widetilde{Y}_{i}=\sum_{j\neq k}|\hat{\mathbf{h}}_{j}^{\text{H}}\mathbf{w}_{i}|^{2}$
converges to \ac{iid} chi-square random variables $\frac{1}{2}\chi^{2}(2(K-1))$.
The limiting \ac{cdf} of $\widetilde{Y}_{i}$ is thus given by $F_{K}(y)=\frac{1}{\Gamma(K-1)}\overline{\gamma}(y;K-1)$,
$y\geq0$, where $\overline{\gamma}(y;k)=\int_{0}^{y}u^{k-1}e^{-u}du$
is the incomplete gamma function. 

Define $F_{K}^{*}(y)=F_{K}(-\frac{1}{y})$ for $y\leq0$. Then the
following property holds 
\begin{align}
 & \lim_{t\to-\infty}\frac{F_{K}^{*}(ty)}{F_{K}^{*}(t)}=\lim_{t\to-\infty}\frac{\overline{\gamma}(-\frac{1}{2ty};K-1)}{\overline{\gamma}(-\frac{1}{2t};K-1)}\nonumber \\
 & \qquad=\lim_{t\to-\infty}\frac{\overline{\gamma}(-\frac{1}{2ty};K-1)}{\left(-\frac{1}{2ty}\right)^{K-1}}\frac{\left(-\frac{1}{2t}\right)^{K-1}}{\overline{\gamma}(-\frac{1}{2t};K-1)}\frac{\left(-\frac{1}{2ty}\right)^{K-1}}{\left(-\frac{1}{2t}\right)^{K-1}}\nonumber \\
 & \qquad=y^{-(K-1)}\label{eq:app-proof-exetrem-value-theorm-cond1}
\end{align}
where we used the property of incomplete gamma function that $\lim_{x\to0}\overline{\gamma}(x;k)/x^{k}=\frac{1}{k}$.

The extreme value theory \cite[Theorem 2.1.5]{Galambos:78book} concludes
that under condition (\ref{eq:app-proof-exetrem-value-theorm-cond1}),
\[
\lim_{N\to\infty}\lim_{N_{t}\to\infty}\mathbb{P}\left\{ \min_{i=1,2,\dots,N}\widetilde{Y}_{i}<\phi_{N}y\right\} =1-\exp(-y^{K-1})
\]
for $y\geq0$, where $\phi_{N}=\sup\{y:F_{K}(y)\leq\frac{1}{N}\}$
which yields (\ref{eq:phi_N}).

Moreover, using the limiting property of the incomplete gamma function
$\overline{\gamma}(y;k)=\frac{y^{k}}{k}+o(y^{k})$, we have $F_{K}(y)\approx\frac{y^{K-1}}{(K-1)\Gamma(K-1)}=\frac{y^{K-1}}{\Gamma(K)}$.
Solving 
\[
F_{K}(y)\approx\frac{y^{K-1}}{\Gamma(K)}=\frac{1}{N}
\]
 for $y$, gives $\phi_{N}\approx\hat{y}=\Gamma(K)^{-\frac{1}{K-1}}N^{-\frac{1}{K-1}}$
as in (\ref{eq:phi_N_approx}).

Note that since the $F_{K}(y)$ decreases when $K$ increases, thus
the optimal solution $y^{*}=F_{K}^{-1}(\frac{1}{N})$ decreases as
$K$ decreases. Meanwhile, the approximation $F_{K}(y)\approx y^{K-1}/\Gamma(K)$
is asymptotically accurate when $y$ approaches $0$. This means that
the approximation of $\phi_{N}$ becomes accurate for small $K$. 


\section{Proof of Lemma \ref{lem:partial-CSI-model}}

\label{sec:app-proof-lem-partial-CSI-model}

Under the \ac{klt} (\ref{eq:klt}), the vector $\mathbf{q}_{k}^{(j)}$
has independent elements, where the $i$th complex element $\big[\mathbf{q}_{k}^{(j)}\big]_{i}$
has variance $\lambda_{kj}^{(i)}.$ According to Shannon's distortion-rate
theory, the minimum distortion of the $i$th complex element is given
by 
\[
D_{kj}^{(i)}\triangleq\mathbb{E}\left\{ \left(\big[\mathbf{q}_{k}^{(j)}\big]_{i}-\big[\hat{\mathbf{q}}_{k}^{(j)}\big]_{i}\right)^{2}\right\} =\lambda_{kj}^{(i)}2^{-b_{kj}^{(i)}}
\]
where $b_{kj}^{(i)}$ is the number of bits allocated to the $i$th
complex element of $\mathbf{q}_{k}^{(j)}$. Therefore, the minimum
distortion $D_{kj}=\sum_{i=1}^{M_{kj}}D_{kj}^{(i)}$ can be achieved
by 
\begin{align}
\underset{\{b_{kj}^{(i)}\geq0\}}{\mbox{minimize}} & \quad D_{kj}=\sum_{i=1}^{M_{kj}}\lambda_{kj}^{(i)}2^{-b_{kj}^{(i)}}\label{eq:minimum-distortion-bit-allocation-problem}\\
\mbox{subject to} & \quad\sum_{i=1}^{M_{kj}}b_{kj}^{(i)}=b_{kj}\nonumber 
\end{align}

\begin{lemma}
With sufficiently large $b_{kj}$, the minimum value of (\ref{eq:minimum-distortion-bit-allocation-problem})
is given by 
\[
D_{kj}^{*}=M_{kj}\Big(\prod_{i=1}^{M_{kj}}\lambda_{kj}^{(i)}\Big)^{\frac{1}{M_{kj}}}2^{-\frac{b_{kj}}{M_{kj}}}
\]
and the distortion of each element $\big[\mathbf{q}_{k}^{(j)}\big]_{i}$
is $D_{kj}^{(i)}=\frac{1}{M_{kj}}D_{kj}^{*}$, for $i=1,2,\dots,M_{kj}$. 
\end{lemma}

\begin{IEEEproof}
The closed-form solution to (\ref{eq:minimum-distortion-bit-allocation-problem})
can be derived using Lagrangian methods. Details are omitted here
due to page limit. 
\end{IEEEproof}

Let $\mathbf{q}_{k}^{(j)}=\tilde{\beta}_{kj}\hat{\mathbf{q}}_{k}^{(j)}+\tilde{\tau}_{kj}\tilde{\mathbf{q}}_{k}^{(j)}$,
where 
\begin{equation}
\tilde{\beta}_{kj}=\hat{\mathbf{q}}_{k}^{(j)\text{H}}\mathbf{q}_{k}^{(j)}/\|\hat{\mathbf{q}}_{k}^{(j)}\|^{2}\label{eq:tau-definition}
\end{equation}
and $\tilde{\mathbf{q}}_{k}^{(j)}$ is normalized to $\|\tilde{\mathbf{q}}_{k}^{(j)}\|=1$.
Note that $\tilde{\mathbf{q}}_{k}^{(j)}$ is orthogonal to $\hat{\mathbf{q}}_{k}^{(j)}$,
because $\tilde{\beta}_{kj}\hat{\mathbf{q}}_{k}^{(j)}$ is orthogonal
projection of\textbf{ }$\mathbf{q}_{k}^{(j)}$ onto $\hat{\mathbf{q}}_{k}^{(j)}$,
and hence the residual $\mathbf{q}_{k}^{(j)}-\tilde{\beta}_{kj}\hat{\mathbf{q}}_{k}^{(j)}$
is orthogonal to $\hat{\mathbf{q}}_{k}^{(j)}$. Since the distortion
of every element of $\mathbf{q}_{k}^{(j)}$ is the same, we have $\mathbf{q}_{k}^{(j)}-\hat{\mathbf{q}}_{k}^{(j)}\sim\mathcal{CN}(\mathbf{0},\frac{D_{kj}^{*}}{M_{kj}}\mathbf{I})$,
and hence $\tilde{\mathbf{q}}_{k}^{(j)}$ is an isotropic random vector
in $M_{kj}$ dimensional subspace. 

To quantify $\tilde{\tau}_{kj}$, we have 
\begin{align*}
D_{kj} & =\mathbb{E}\Big\{\big\|\tilde{\beta}_{kj}\hat{\mathbf{q}}_{k}^{(j)}+\tilde{\tau}_{kj}\tilde{\mathbf{q}}_{k}^{(j)}-\hat{\mathbf{q}}_{k}^{(j)}\big\|^{2}\Big\}\\
 & =\mathbb{E}\Big\{\big\|(\tilde{\beta}_{kj}-1)\hat{\mathbf{q}}_{k}^{(j)}+\tilde{\tau}_{kj}\tilde{\mathbf{q}}_{k}^{(j)}\big\|^{2}\Big\}\\
 & =\mathbb{E}\Big\{\big\|(\tilde{\beta}_{kj}-1)\hat{\mathbf{q}}_{k}^{(j)}\big\|^{2}\Big\}+\mathbb{E}\Big\{\tilde{\tau}_{kj}^{2}\Big\}\\
 & \geq\mathbb{E}\Big\{\tilde{\tau}_{kj}^{2}\Big\}
\end{align*}
where the third equality is due to the orthogonality between $\hat{\mathbf{q}}_{k}^{(j)}$
and $\widetilde{\mathbf{q}}_{k}^{(j)}$.

As a result, we have 
\[
\mathbb{E}\Big\{\tilde{\tau}_{kj}^{2}\Big\}\leq D_{kj}^{*}=M_{kj}\bigg(\prod_{i=1}^{M_{kj}}\lambda_{kj}^{(i)}\bigg)^{\frac{1}{M_{kj}}}2^{-\frac{b_{kj}}{M_{kj}}}
\]
for large $b_{kj}$. Moreover, 
\begin{align*}
\mathbf{h}_{k}^{(j)} & =\sqrt{l_{k}}\mathbf{U}_{j}\mathbf{U}_{kj}\mathbf{q}_{k}^{(j)}\\
 & =\sqrt{l_{k}}\mathbf{U}_{j}\mathbf{U}_{kj}\tilde{\beta}_{kj}\hat{\mathbf{q}}_{k}^{(j)}+\sqrt{l_{k}}\mathbf{U}_{j}\mathbf{U}_{kj}\tilde{\tau}_{kj}\tilde{\mathbf{q}}_{k}^{(j)}\\
 & =\beta_{kj}\mathbf{U}_{j}\hat{\mathbf{g}}_{k}^{(j)}+\tau_{kj}\mathbf{U}_{j}\mathbf{s}_{k}^{(j)}
\end{align*}
where $\tau_{kj}\triangleq\sqrt{l_{k}}\tilde{\tau}_{kj}$, 
\begin{align*}
\beta_{kj}\triangleq\sqrt{l_{k}}\tilde{\beta}_{kj} & =\sqrt{l_{k}}\hat{\mathbf{q}}_{k}^{(j)\text{H}}\mathbf{U}_{kj}^{\text{H}}\mathbf{U}_{kj}\mathbf{q}_{k}^{(j)}/\|\mathbf{U}_{kj}\hat{\mathbf{q}}_{k}^{(j)}\|^{2}\\
 & =\sqrt{l_{k}}\hat{\mathbf{g}}_{k}^{(j)\text{H}}\mathbf{g}_{k}^{(j)}/\|\hat{\mathbf{g}}_{k}^{(j)}\|^{2}
\end{align*}
and $\mathbf{s}_{k}^{(j)}=\mathbf{U}_{kj}\tilde{\mathbf{q}}_{k}^{(j)}$
is an isotropic unit norm random vector orthogonal to $\hat{\mathbf{g}}_{k}^{(j)}$,
since$\mathbf{U}_{kj}$ is a unitary matrix.

\section{Proof of Lemma \ref{lem:Virtual-SLNR-Lower-bound}}

\label{sec:app-proof-lem-virtual-slnr-lower-bound}


The virtual SLNR $\bar{\Gamma}_{k}$ can be lower bounded by 
\begin{align}
\bar{\Gamma}_{k} & =\mathbb{E}\left\{ \frac{|\mathbf{h}_{k}^{\text{H}}\mathbf{w}_{k}^{\text{SLNR}}|^{2}}{\sum_{j\neq k}|\mathbf{h}_{j}^{\text{H}}\mathbf{w}_{k}^{\text{SLNR}}|^{2}+\alpha}\right\} \nonumber \\
 & \stackrel{(a)}{\geq}\mathbb{E}\left\{ \frac{|\mathbf{h}_{k}^{\text{H}}\mathbf{w}_{k}^{\text{ZF}}|^{2}}{\sum_{j\neq k}|\mathbf{h}_{j}^{\text{H}}\mathbf{w}_{k}^{\text{ZF}}|^{2}+\alpha}\right\} \nonumber \\
 & \stackrel{(b)}{=}\mathbb{E}\bigg\{\mathbb{E}\left\{ |\mathbf{h}_{k}^{\text{H}}\mathbf{w}_{k}^{\text{ZF}}|^{2}\Big|\mathbf{w}_{k}^{\text{ZF}}\right\} \mathbb{E}\Big\{\frac{1}{\sum_{j\neq k}|\mathbf{h}_{j}^{\text{H}}\mathbf{w}_{k}^{\text{ZF}}|^{2}+\alpha}\Big|\mathbf{w}_{k}^{\text{ZF}}\Big\}\bigg\}\nonumber \\
 & \stackrel{(c)}{\geq}\mathbb{E}\Bigg\{\frac{\mathbb{E}\left\{ |\mathbf{h}_{k}^{\text{H}}\mathbf{w}_{k}^{\text{ZF}}|^{2}\big|\mathbf{w}_{k}^{\text{ZF}}\right\} }{\sum_{j\neq k}\mathbb{E}\Big\{|\mathbf{h}_{j}^{\text{H}}\mathbf{w}_{k}^{\text{ZF}}|^{2}\big|\mathbf{w}_{k}^{\text{ZF}}\Big\}+\alpha}\Bigg\}\label{eq:app-Gamma-lower-bound-1}
\end{align}
where $\mathbf{w}_{k}^{\text{SLNR}}$ is the SLNR precoder in continuous
domain given in (\ref{eq:w-slnr}), $\mathbf{w}_{k}^{\text{ZF}}=\tilde{\mathbf{w}}_{k}^{\text{ZF}}/\|\tilde{\mathbf{w}}_{k}^{\text{ZF}}\|$
and $\tilde{\mathbf{w}}_{k}^{\text{ZF}}$ is the $k$th column of
the \ac{zf} precoding matrix $\tilde{\mathbf{W}}_{k}=\hat{\mathbf{H}}_{k}(\hat{\mathbf{H}}_{k}^{\text{H}}\hat{\mathbf{H}}_{k})^{-1}$.
Inequality $\stackrel{(a)}{\geq}$ is due to the fact that $\mathbf{w}_{k}^{\text{ZF}}$
is not optimal in maximizing the SLNR criterion (\ref{eq:w-slnr}).
Equality $\stackrel{(b)}{=}$ is because $|\mathbf{h}_{k}^{\text{H}}\mathbf{w}_{k}^{\text{ZF}}|^{2}$
is independent to $|\mathbf{h}_{j}^{\text{H}}\mathbf{w}_{k}^{\text{ZF}}|^{2}$
given the precoder $\mathbf{w}_{k}^{\text{ZF}}$, and the outer expectation
$\mathbb{E}\{\cdot\}$ is taken over the randomness of $\mathbf{w}_{k}^{\text{ZF}}$.
Furthermore, inequality $\stackrel{(c)}{\geq}$ is from the Jesen's
inequality $\mathbb{E}\{f(x)\}\geq f(\mathbb{E}\{x\})$ for the convex
function $f(x)=1/(x+\alpha)$. 

\subsection{The Interference Term}

Using the partial CSI model (\ref{eq:partial-CSI-model}) and note
that $\mathbf{h}_{j}=\mathbf{h}_{j}^{(k)}+\mathbf{h}_{j}^{(k)\perp}$,
the interference term in the denominator of (\ref{eq:app-Gamma-lower-bound-1})
can be derived as 
\begin{align}
 & \mathbb{E}\Big\{|\mathbf{h}_{j}^{\text{H}}\mathbf{w}_{k}^{\text{ZF}}|^{2}\big|\mathbf{w}_{k}^{\text{ZF}}\Big\}\nonumber \\
 & \quad=\mathbb{E}_{\mathbf{w}}\bigg\{\bigg|\left(\beta_{jk}\mathbf{U}_{k}\hat{\mathbf{g}}_{j}^{(k)}+\tau_{jk}\mathbf{U}_{k}\mathbf{s}_{j}^{(k)}+\mathbf{h}_{j}^{(k)\perp}\right)^{\text{H}}\mathbf{w}_{k}^{\text{ZF}}\bigg|^{2}\bigg\}\nonumber \\
 & \quad=\mathbb{E}_{\mathbf{w}}\left\{ \big|\tau_{jk}\mathbf{s}_{j}^{(k)\text{H}}\mathbf{U}_{k}^{\text{H}}\mathbf{w}_{k}^{\text{ZF}}\big|^{2}\right\} \label{eq:Ijk-1}
\end{align}
where $\mathbb{E}_{\mathbf{w}}\{\cdot\}\triangleq\mathbb{E}\{\cdot\big|\mathbf{w}_{k}^{\text{ZF}}\}$
is used here as the shorthand notation of the expectation conditioned
on $\mathbf{w}_{k}^{\text{ZF}}$. Note that $\mathbf{w}_{k}^{\text{ZF}}$
is orthogonal to $\hat{\mathbf{g}}_{j}^{(k)}$ and $\mathbf{h}_{j}^{(k)\perp}$,
as $\mathbf{h}_{j}^{(k)\perp}$ lies in the orthogonal subspace of
$\mathbf{h}_{k}$. 

Let $\mathbf{v}_{j}^{(k)}=\mathbf{U}_{k}^{\text{H}}\mathbf{w}_{k}^{\text{ZF}}/\|\mathbf{U}_{k}^{\text{H}}\mathbf{w}_{k}^{\text{ZF}}\|$.
Note that, $\mathbf{s}_{j}^{(k)}$ is an isotropic random vector independent
to $\mathbf{v}_{j}^{(k)}$. As a result, $|\mathbf{s}_{j}^{(k)\text{H}}\mathbf{v}_{j}^{(k)}|^{2}$
follows beta distribution $\mathcal{B}(1,M_{jk}-1)$ with parameters
$1$ and $M_{jk}-1$ according to the following lemma. 
\begin{lemma}
[Isotropic Vectors \cite{au2007performance}]\label{lem:Isotropic-Vectors}
Let $\mathbf{u},\mathbf{v}\in\mathbb{C}^{M}$ be two random vectors
that follow distribution $\mathcal{C}\mathcal{N}(0,\sigma^{2}\mathbf{I})$.
Let $\widetilde{\mathbf{u}}=\mathbf{u}/\|\mathbf{u}\|$ and $\widetilde{\mathbf{v}}=\mathbf{v}/\|\mathbf{v}\|$.
Then the quantity $|\widetilde{\mathbf{u}}^{\text{H}}\widetilde{\mathbf{v}}|^{2}$
follows beta distribution $\mathcal{B}(1,M-1)$ with parameters $1$
and $M-1$. 
\end{lemma}

The interference term in (\ref{eq:Ijk-1}) is thus bounded as 
\begin{align*}
\mathbb{E}_{\mathbf{w}}\Big\{|\mathbf{h}_{j}^{\text{H}}\mathbf{w}_{k}^{\text{ZF}}|^{2}\Big\} & =\mathbb{E}_{\mathbf{w}}\left\{ \Big|\tau_{jk}\mathbf{s}_{j}^{(k)\text{H}}\mathbf{v}_{j}^{(k)}\|\mathbf{U}_{k}^{\text{H}}\mathbf{w}_{k}^{\text{ZF}}\|\Big|^{2}\right\} \\
 & \stackrel{(a)}{\leq}\mathbb{E}_{\mathbf{w}}\Big\{\tau_{jk}^{2}\mathcal{B}(1,M_{jk}-1)\|\mathbf{U}_{k}\|^{2}\|\mathbf{w}_{k}^{\text{ZF}}\|^{2}\Big\}\\
 & =\mathbb{E}_{\mathbf{w}}\Big\{\tau_{jk}^{2}\mathcal{B}(1,M_{jk}-1)\Big\}\\
 & \stackrel{(b)}{=}\mathbb{E}\{\tau_{jk}^{2}\}\mathbb{E}\{\mathcal{B}(1,M_{jk}-1)\}\\
 & \leq l_{j}\bigg(\prod_{i=1}^{M_{jk}}\lambda_{jk}^{(i)}\bigg)^{\frac{1}{M_{jk}}}2^{-\frac{b_{jk}}{M_{jk}}}
\end{align*}
where $\stackrel{(a)}{\leq}$ is from triangle inequality $\Big|\mathbf{s}_{j}^{(k)\text{H}}\mathbf{v}_{j}^{(k)}\|\mathbf{U}_{k}^{\text{H}}\mathbf{w}_{k}^{\text{ZF}}\|\Big|^{2}\leq|\mathbf{s}_{j}^{(k)\text{H}}\mathbf{v}_{j}^{(k)}|^{2}\|\mathbf{U}_{k}\|^{2}\|\mathbf{w}_{k}^{\text{ZF}}\|^{2}$,
$\stackrel{(b)}{=}$ is from the fact that $\tau_{jk}^{2}\mathcal{B}(1,M_{jk}-1)$
is independent to $\mathbf{w}_{k}^{\text{ZF}}$, and in addition,
$\mathbb{E}\{\mathcal{B}(1,M_{jk}-1)\}=1/M_{jk}$. 

\subsection{The signal term}

The signal term $\mathbb{E}\left\{ |\mathbf{h}_{k}^{\text{H}}\mathbf{w}_{k}^{\text{ZF}}|^{2}\right\} $
can be computed as follows.

Let 
\[
\mathbf{P}_{k}=\mathbf{I}-\hat{\mathbf{H}}_{-k}\big(\hat{\mathbf{H}}_{-k}^{\text{H}}\hat{\mathbf{H}}_{-k}\big)^{-1}\hat{\mathbf{H}}_{-k}^{\text{H}}
\]
 be a $N_{t}\times N_{t}$ projection matrix for user $k$, where
$\hat{\mathbf{H}}_{-k}=\Big[\big\{\hat{\mathbf{h}}_{j}^{(k)}:j\neq k\big\}\Big]$
is a $N_{t}\times(K-1)$ CSI matrix that contains the CSI exchanged
from all the other users. As a result, the ZF precoder $\mathbf{w}_{k}^{\text{ZF}}$
can be equivalently written as 
\[
\mathbf{w}_{k}^{\text{ZF}}=\frac{\mathbf{P}_{k}\mathbf{h}_{k}}{\|\mathbf{P}_{k}\mathbf{h}_{k}\|}.
\]

Using the property of a projection matrix $\mathbf{P}_{k}=\mathbf{P}_{k}\mathbf{P}_{k}^{\text{H}}=\mathbf{P}_{k}^{\text{H}}$,
the following holds 
\[
|\mathbf{h}_{k}^{\text{H}}\mathbf{w}_{k}^{\text{ZF}}|^{2}=\frac{|\mathbf{h}_{k}^{\text{H}}\mathbf{P}_{k}\mathbf{h}_{k}|^{2}}{\|\mathbf{P}_{k}\mathbf{h}_{k}\|^{2}}=\frac{\|\mathbf{h}_{k}^{\text{H}}\mathbf{P}_{k}^{\text{H}}\mathbf{P}_{k}\mathbf{h}_{k}\|^{2}}{\|\mathbf{P}_{k}\mathbf{h}_{k}\|^{2}}=\|\mathbf{P}_{k}\mathbf{h}_{k}\|^{2}.
\]
As a result, 
\begin{align*}
\mathbb{E}\left\{ |\mathbf{h}_{k}^{\text{H}}\mathbf{w}_{k}^{\text{ZF}}|^{2}\Big|\hat{\mathbf{H}}_{-k}\right\}  & =\mathbb{E}\left\{ \|\mathbf{P}_{k}\mathbf{h}_{k}\|^{2}\Big|\hat{\mathbf{H}}_{-k}\right\} \\
 & =\mathbb{E}\left\{ \mbox{tr}\big\{\mathbf{P}_{k}\mathbf{h}_{k}\mathbf{h}_{k}^{\text{H}}\mathbf{P}_{k}^{\text{H}}\big\}\Big|\hat{\mathbf{H}}_{-k}\right\} \\
 & \stackrel{(a)}{=}\mbox{tr}\left\{ \mathbf{P}_{k}\mathbb{E}\Big\{\mathbf{h}_{k}\mathbf{h}_{k}^{\text{H}}\Big|\hat{\mathbf{H}}_{-k}\Big\}\mathbf{P}_{k}^{\text{H}}\right\} \\
 & \stackrel{(b)}{=}\mbox{tr}\left\{ \mathbf{P}_{k}l_{k}\mathbf{R}_{k}\mathbf{P}_{k}^{\text{H}}\right\} \\
 & \stackrel{(c)}{\geq}l_{k}\sum_{i=K}^{N_{t}}\lambda_{k}^{(i)}
\end{align*}
where $\lambda_{k}^{(i)}$ are the eigenvalues of $\mathbf{R}_{k}$
in descending order, the equality $\stackrel{(a)}{=}$ is because
$\mathbf{P}_{k}$ only depends on $\hat{\mathbf{H}}_{-k}$, the equality
$\stackrel{(b)}{=}$ is due to the independence between $\mathbf{h}_{k}$
and $\hat{\mathbf{H}}_{-k}$, and the lower bound $\stackrel{(c)}{\geq}$
is tight when $\mathbf{P}_{k}$ is to project $\mathbf{R}_{k}$ onto
the orthogonal subspace of the subspace that is spanned by the $K-1$
dominant eigenvectors of $\mathbf{R}_{k}$. 

Therefore, 
\[
\mathbb{E}\left\{ |\mathbf{h}_{k}^{\text{H}}\mathbf{w}_{k}^{\text{ZF}}|^{2}\right\} =\mathbb{E}\left\{ \mathbb{E}\left\{ |\mathbf{h}_{k}^{\text{H}}\mathbf{w}_{k}^{\text{ZF}}|^{2}\Big|\hat{\mathbf{H}}_{-k}\right\} \right\} \geq l_{k}\sum_{i=K}^{N_{t}}\lambda_{k}^{(i)}.
\]

\subsection{The Lower Bound}

The virtual SLNR can be further bounded as 
\begin{align*}
\bar{\Gamma}_{k} & \geq\mathbb{E}\Bigg\{\frac{\mathbb{E}\left\{ |\mathbf{h}_{k}^{\text{H}}\mathbf{w}_{k}^{\text{ZF}}|^{2}\big|\mathbf{w}_{k}^{\text{ZF}}\right\} }{\sum_{j\neq k}l_{j}\bigg(\prod_{i=1}^{M_{jk}}\lambda_{jk}^{(i)}\bigg)^{\frac{1}{M_{jk}}}2^{-\frac{b_{jk}}{M_{jk}}}+\alpha}\Bigg\}\\
 & \geq\frac{l_{k}\sum_{i=K}^{N_{t}}\lambda_{k}^{(i)}}{\sum_{j\neq k}l_{j}\bigg(\prod_{i=1}^{M_{jk}}\lambda_{jk}^{(i)}\bigg)^{\frac{1}{M_{jk}}}2^{-\frac{b_{jk}}{M_{jk}}}+\alpha}
\end{align*}
which proves the result.

\section{Proof of Theorem \ref{thm:Optimal-Bit-Partition}}

\label{sec:app-proof-thm-optimal-bit-partition}

As the constrained minimization problem (\ref{eq:bit-partition-problem-minimization})
is convex, it can be solved using Lagrangian methods. Specifically,
the Lagrangian function of (\ref{eq:bit-partition-problem-minimization})
can be written as 
\[
\mathcal{L}(\mathbf{b},\mu)=\sum_{k=1}^{K}\log\Bigg(\sum_{j\neq k}\omega_{jk}2^{-\frac{b_{jk}}{M_{jk}}}+\alpha\Bigg)+\mu\bigg(\sum_{k=1}^{K}\sum_{j\neq k}b_{kj}-B_{\text{tot}}\bigg)
\]
and the \ac{kkt} condition is given by 
\begin{align}
 & \frac{\partial\mathcal{L}(\mathbf{b},\mu)}{\partial b_{jk}}=\frac{-\frac{\ln2}{M_{jk}}\omega_{jk}2^{-\frac{b_{jk}}{M_{jk}}}}{\sum_{m\neq k}\omega_{mk}2^{-\frac{b_{mk}}{M_{mk}}}+\alpha}+\mu=0,\;b_{jk}\geq0\label{eq:kkt-condition-1}\\
 & \qquad\qquad\qquad\qquad\qquad\qquad\forall j\neq k,\;k=1,2,\dots,K\nonumber \\
 & \mu\bigg(\sum_{k=1}^{K}\sum_{j\neq k}b_{kj}-B_{\text{tot}}\bigg)=0,\qquad\mu\geq0.\label{eq:kkt-condition-2}
\end{align}

Condition (\ref{eq:kkt-condition-1}) can be divided into $K$ sets
of equations. Each set consists of $K-1$ equations as follows
\[
\frac{\ln2}{M_{jk}}\omega_{jk}2^{-\frac{b_{jk}}{M_{jk}}}-\mu\sum_{m\neq k}\omega_{mk}2^{-\frac{b_{mk}}{M_{mk}}}=\mu\alpha,\qquad j\neq k
\]
which can be written into a compact form as 
\[
\mathbf{A}_{k}\mathbf{x}_{k}-\mu\mathbf{1}\bm{\omega}_{k}^{\text{T}}\mathbf{x}_{k}=\mu\alpha
\]
where 
\[
\mathbf{x}_{k}=(2^{-\frac{b_{1k}}{M_{1k}}},2^{-\frac{b_{2k}}{M_{2k}}},\dots,2^{-\frac{b_{k-1,k}}{M_{k-1,k}}},2^{-\frac{b_{k+1,k}}{M_{k+1,k}}},\dots,2^{-\frac{b_{Kk}}{M_{Kk}}})^{\text{T}}.
\]
This leads to solutions (\ref{eq:optimal-bit-partition-solution})
and (\ref{eq:optimal-bit-partition-solution-auxilliary}), where the
projection $[\cdot]^{+}$ and the choice of $\mu$ are to satisfy
the \ac{kkt} conditions (\ref{eq:optimal-bit-partition-solution})
and (\ref{eq:kkt-condition-2}).

%


%

\bibliographystyle{IEEEtran}
\bibliography{IEEEabrv,/Users/Allen/Dropbox/Draft/Bibliography/StringDefinitions,/Users/Allen/Dropbox/Draft/My_reference}

\end{document}